# Dynamics of frequency-swept nuclear spin optical pumping in powdered diamond at low magnetic fields

Pablo R. Zangara[1], Siddharth Dhomkar[1], Ashok Ajoy[3], Kristina Liu[3], Raffi Nazaryan[3], Daniela Pagliero[1], Dieter Suter[6], Jeffrey A. Reimer[4,5], Alexander Pines[3,5], Carlos A. Meriles[1,2]

[1]Deptartment of Physics, City College of New York, City University of New York, New York, New York 10031, USA. [2]Graduate Center, City University of New York, New York, New York 10016, USA. [3]Department of Chemistry, University of California at Berkeley, Berkeley, California 94720, USA. [4]Department of Chemical and Biomolecular Engineering, University of California at Berkeley, Berkeley, California 94720, USA. [5]Materials Science Division, Lawrence Berkeley National Laboratory, Berkeley, California 94720, USA. [6]Fakultät Physik, Technische Universität Dortmund, D-44221 Dortmund, Germany.

**A broad effort is underway to improve the sensitivity of nuclear magnetic resonance through the use of dynamic nuclear polarization. Nitrogen-vacancy (NV) centers in diamond offer an appealing platform because these paramagnetic defects can be optically polarized efficiently at room temperature. However, work thus far has been mainly limited to single crystals because most polarization transfer protocols are sensitive to misalignment between the NV and magnetic field axes. Here we study the spin dynamics of NV-$^{13}$C pairs in the simultaneous presence of optical excitation and microwave frequency sweeps at low magnetic fields. We show that a subtle interplay between illumination intensity, frequency sweep rate, and hyperfine coupling strength leads to efficient, sweep-direction-dependent $^{13}$C spin polarization over a broad range of orientations of the magnetic field. In particular, our results strongly suggest that finely-tuned, moderately coupled nuclear spins are key to the hyperpolarization process, which makes this mechanism distinct from other known dynamic polarization channels. These findings pave the route to applications where powders are intrinsically advantageous, including the hyper-polarization of target fluids in contact with the diamond surface or the use of hyperpolarized particles as contrast agents for in-vivo imaging.**

**Nitrogen-vacancy center | hyperpolarization | diamond powder | optical spin pumping | Landau-Zener crossings**

Nuclear magnetic resonance (NMR) has proven to be a powerful tool in areas ranging from molecular analysis to biomedical imaging. Unfortunately, the attainable nuclear spin polarization is often a small fraction of the possible maximum, thus imposing strict constraints on the minimum sample size and acquisition time. Dynamic nuclear polarization (DNP), i.e, the transfer of magnetization from electron to nuclear spins (1), is a route of growing popularity that substantially mitigates this problem. Enhanced polarization can be attained, e.g., with the aid of dissolved molecular radicals, though the most efficient implementations often rely on freeze-thaw protocols and high-frequency microwave (MW) excitation, which are expensive and technically demanding (2).

Adding to the library of DNP platforms, optically active spin-defects in semiconductors are attracting widespread attention as alternative hyperpolarization agents. Among them, the negatively-charged nitrogen vacancy center (NV) in diamond is arguably one of the most promising candidates, since it can be spin-polarized optically to a high degree with only modest illumination intensities and under ambient conditions (3). A variety of protocols have already been implemented to transfer NV spin polarization to surrounding nuclear spins including level-anti-crossing-mediated transfer in the NV$^-$ ground (4) and excited states (5), cross-relaxation with P1-centers (6-8), spin-swap and population trapping (9), amplitude-matched microwave excitation (10,11), and transfer via microwave sweeps (12,13). Despite this progress, however, efficient hyperpolarization of randomly oriented samples at arbitrary magnetic fields has remained elusive, hence precluding applications where the use of diamond powders is desirable or necessary. Examples worth highlighting include the use of particles as contrast agents for in-vivo magnetic resonance imaging (of interest given the biocompatibility of diamond (14)) or as a source of nuclear spin polarization in fluids (attractive given the enhanced surface-to-volume ratio inherent to powders).

Recent work demonstrated efficient $^{13}$C DNP in diamond powders simultaneously exposed to optical illumination and microwave (MW) frequency sweeps (15), but gaining a detailed understanding of the microscopic mechanisms at play has proven subtle due to a complex interplay between the multiple degrees of freedom. Here we examine the dynamics of an NV-$^{13}$C spin pair undergoing simultaneous optical illumination and MW excitation. We focus on the limit of low magnetic fields (~10-30 mT) and consider the system evolution in the presence of MW sweeps of variable sweep rate. Through a transformation to the rotating frame, we show that the dynamics can be described in terms of a series of multi-branched Landau-Zener crossings; the branch-dependent degree of adiabaticity during these crossings combined with mild optical pumping of the NV spin leads to a net buildup of $^{13}$C polarization, which is robust against NV misalignment and efficient for hyperfine couplings as low as 0.2-0.3 MHz. In particular, we show that moderately-coupled carbons are dominant in driving the polarization dynamics of the bulk, a feature very much in contrast with prior spin transfer studies in diamond (mediated by first or second shell carbons). For the present experimental conditions, the observed level of $^{13}$C polarization is in the range 0.1-0.3%, corresponding to a one- to three-hundred-fold enhancement over the thermal polarization at 7 T. These results can be immediately extended to paramagnetic defects other than the NV (such as the neutrally charged silicon-vacancy (SiV) center (16,17)) or to other wide-bandgap semiconductors (such as silicon-carbide) hosting point defects that can be optically spin polarized (18).

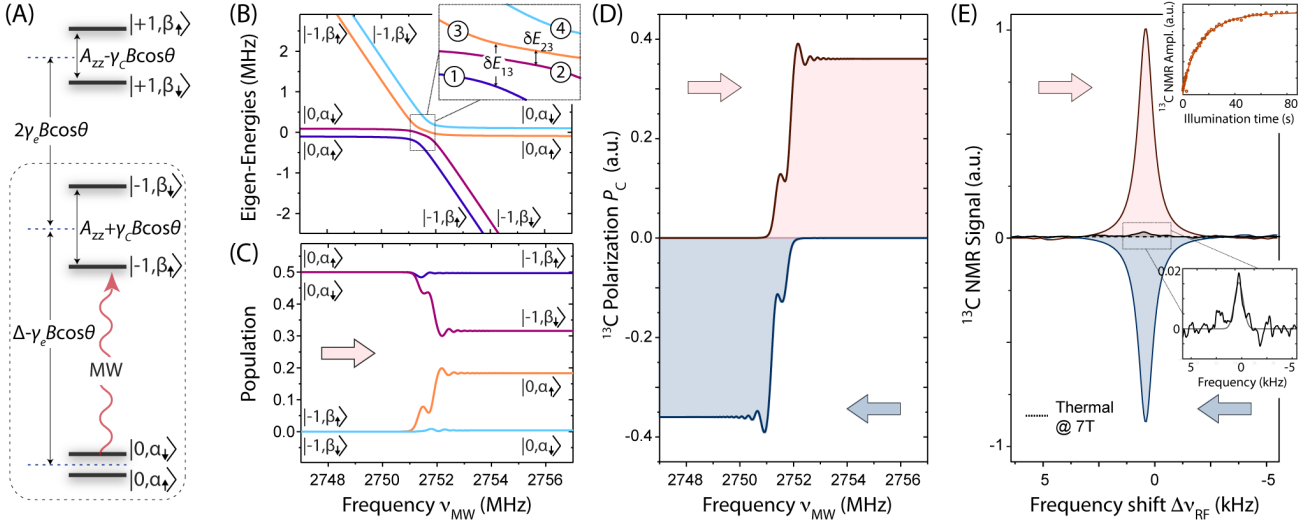

**Fig. 1**. Dynamic polarization of a $^{13}$C nuclear spin coupled to an NV center. (A) Schematics of the ground state energy diagram of an NV–$^{13}$C pair. In each ket, the first (second) index refers to the electron (nuclear) spin quantum number, and we assume $A_{zz}$ is positive; all energy separations are approximate. (B) Four lowest eigen-energies of $H_{eff}$ as a function of the MW frequency $\nu_{MW}$ in the region near the $m_s = 0 \leftrightarrow m_s = -1$ transition. The labels denote the corresponding crystal-frame eigenstates. (C) Population of the instantaneous eigenstates for the eigenenergies shown in (B) as the system undergoes a MW sweep at a velocity $\dot{\nu}$=40 MHz/ms. (D) Calculated $^{13}$C polarization during the sweep for low-to-high and high-to-low sweeps (top and bottom plots, respectively). (E) $^{13}$C NMR spectrum at 7 T from single crystal diamond upon 10 s of 532 nm, 1 W laser illumination and MW excitation at 38 mT followed by sample shuttling. The red (blue) trace corresponds to a low-to-high (high-to-low) MW frequency sweep after 20 repetitions of the polarization protocol; in both cases the sweep rate is $\dot{\nu}$ = 60 MHz/ms and the frequency range is ~0.37 GHz, from 3.507 GHz to 3.878 GHz (corresponding to a repetition rate of 164 Hz). The black trace is a reference spectrum from thermal polarization at 7 T for a total of 120 repeats separated by a 600 s wait interval; the attained enhancement relative to the 7 T signal is ~300, corresponding to a bulk $^{13}$C polarization of ~0.3%. In these NMR spectra, $\Delta\nu_{RF} \equiv \nu_C^{(0)} - \nu_{RF}$ is the radio-frequency (RF) shift relative to the high-field $^{13}$C resonant frequency $\nu_C^{(0)} = 70$ MHz. The upper insert shows the NMR peak amplitude as a function of the illumination time; the lower insert is a blowout of the thermal $^{13}$C NMR spectrum. In (B-D) we consider $A_{zz} = A_{xx} = A_{yy} = 500$ kHz, $A_{zx} = 0.3A_{zz}$, $\Omega = 250$ kHz, $B = 10$ mT, and $\theta = 65°, \phi = 0°$; in (C, D) we assume that the electron spin has been fully initialized into $m_S = 0$.

**Results and discussion**

Let us first consider an NV center spin interacting with a single $^{13}$C nucleus in the presence of an external magnetic field,

$$H = \Delta((S^z)^2 - S(S+1)/3) - \gamma_e \mathbf{B} \cdot \mathbf{S} - \gamma_C \mathbf{B} \cdot \mathbf{I} + \mathbf{S} \cdot \mathbf{A} \cdot \mathbf{I}. \quad (1)$$

Here, $\gamma_e$ and $\gamma_C$ stand for the electron and $^{13}$C gyromagnetic ratios, $\mathbf{B}$ is the external magnetic field, and $\mathbf{A}$ is the hyperfine coupling tensor between the NV electronic spin $\mathbf{S}$ and the $^{13}$C spin $\mathbf{I}$. Unless explicitly stated, we consider $|\mathbf{B}| = 10$ mT and 100 kHz $\lesssim \|\mathbf{A}\| \lesssim 10$ MHz, meaning that the characteristic energy scales can be ordered as

$$\Delta \sim 3 \times 10^3 \text{ MHz} > \gamma_e B \sim 3 \times 10^2 \text{ MHz} > \\ > \|\mathbf{A}\| > \gamma_C B \sim 1 \times 10^{-1} \text{ MHz}. \quad (2)$$

In this regime, the crystal field $\Delta$ is dominant and defines the main quantization direction, here chosen to coincide with the crystal frame $z$-axis; accordingly, we write the magnetic field as $\mathbf{B} = B(\sin\theta \cos\phi, \sin\theta \sin\phi, \cos\theta)$ where $\theta$ and $\phi$ respectively denote the polar and azimuthal angles, changing randomly from one particle to another in the diamond powder. Further, the magnitude of the hyperfine coupling is greater than the nuclear Zeeman interaction, which makes the present regime different from that governing the integrated solid effect (ISE) at high fields (19).

Extending the analysis above to identify the terms governing the nuclear spin dynamics is considerably more involved because, at the low magnetic fields considered herein, the hyperfine interaction can be dominant. An effective secular approximation valid for any choice of angles $(\theta, \phi)$ can be derived using Average Hamiltonian Theory (AHT) (20,21). Without loss of generality, we assume the carbon atom lies within the $zx$-plane, and obtain (see SI, Section I),

$$H_{sec} = \Delta((S^z)^2 - S(S+1)/3) - \gamma_e B \cos\theta\, S^z \\ - \gamma_C \mathbf{B} \cdot \mathbf{I} + A_{zz} S^z I^z + A_{zx} S^z I^x \\ - \frac{\gamma_e B \sin\theta}{\Delta} \mathbb{M} \otimes \left[ \cos\phi\, (A_{xx} I^x + A_{zx} I^z) + \sin\phi\, A_{yy} I^y \right] \quad (3)$$

where $\mathbb{M}$ is a constant matrix defined in the Hilbert space $\{|m_S = +1\rangle, |m_S = 0\rangle, |m_S = -1\rangle\}$ of the NV,

$$\mathbb{M} = \begin{bmatrix} 1 & 0 & 0 \\ 0 & -2 & 0 \\ 0 & 0 & 1 \end{bmatrix}.$$

Fig. 1A shows a schematic representation of the NV–$^{13}$C energy diagram: For the nuclear spin states, we use the notation $|\alpha_{\uparrow,\downarrow}\rangle$ and $|\beta_{\uparrow,\downarrow}\rangle$ to underscore the difference with the Zeeman basis states $|\uparrow,\downarrow\rangle$ (even if they retain part of their character, see SI, Section II). While the impact of the hyperfine field on the nuclear spin states in the $m_S = \pm 1$ subspaces is well documented (22), misalignment between the NV axis and the external magnetic field — unavoidable in a powdered sample — makes it necessary to take into account an additional contribution — last term in Eq. (3) — active even when $m_S = 0$ (23,24). To illustrate its importance we consider, for example, a hyperfine interaction $\|\mathbf{A}\| \sim 1$MHz, leading to matrix elements of order $\delta \sim \gamma_e B \|\mathbf{A}\|/\Delta \sim 1\times 10^{-1}$MHz. In the $m_S = 0$ subspace, this contribution — which does not commute with $-\gamma_C B$ — is comparable to the Zeeman splitting of the $^{13}$C states (Eq. (2)) and hence cannot be disregarded. Note that this same term is less important in the $m_S = \pm 1$ subspaces (where the fourth and fifth terms in Eq. (3) become non-zero), since the factor scaling down the hyperfine coupling, $\gamma_e B/\Delta$ amounts to only ~0.1 for a 10 mT magnetic field.

Building on Eq. (3), we can now extend our description to include the effect of MW driving: The MW field is modeled by a term $\propto S^x \cos(\omega t)$, and, upon a unitary transformation into the frame rotating at the MW angular frequency $\omega$ (see SI, Section I), we write the final effective Hamiltonian as

$$H_{\text{eff}} = \Delta((S^z)^2 - S(S+1)/3) - \gamma_e B \cos\theta\, S^z + \Omega S^x$$
$$-\omega(S^z)^2 - \gamma_C \mathbf{B}\cdot\mathbf{I} + A_{zz}S^z I^z + A_{zx}S^z I^x$$
$$-\frac{\gamma_e B \sin\theta}{\Delta}\mathbb{M}\otimes\left[\cos\phi\,(A_{xx}I^x + A_{zx}I^z) + \sin\phi\, A_{yy}I^y\right], \quad (4)$$

where we have introduced the Rabi frequency $\Omega$. To illustrate the mechanism of spin polarization in the presence of a MW sweep, we first determine the eigenenergies of $H_{\text{eff}}$ as a function of the MW frequency $\nu_{\text{MW}} = \omega/(2\pi)$ (Fig. 1B), and subsequently calculate the system evolution assuming initialization into $m_S = 0$ (Fig. 1C) for the case of a positive hyperfine coupling (see below). As we tune the MW frequency in and out of resonance — in the present example, from lower to higher frequencies across the $m_S = 0 \leftrightarrow m_S = -1$ subset of transitions — the dynamics that follows can be interpreted in terms of a Landau-Zener (LZ) population exchange near the avoided crossings. The corresponding energy gaps can be derived via second order perturbation theory for the inter-level transitions in Eq. (4); using numeric labels 1 through 4 to identify branches in order of increasing energy (see insert to Fig. 1B), we find (see SI, Section II)

$$\delta E_{13} \sim \Omega, \quad (5)$$

and

$$\delta E_{23} \approx \left| \frac{\omega_{0I}}{2} + \frac{(A_{zx}^2 - 2\Omega^2)}{8(\omega_{0I}+A_{zz})} + \right.$$
$$\left. - \frac{1}{2}\sqrt{\left(\omega_{0I} - \frac{A_{zx}^2}{4(\omega_{0I}+A_{zz})}\right)^2 + \Omega^2} \right|. \quad (6)$$

Net nuclear spin polarization emerges from the nuclear-spin-selective adiabaticity of the MW sweep. Assuming for concreteness a low-to-high-frequency sweep, the gap $\delta E_{13}$ yields an LZ jump probability between branches 1 and 3

$$p(1|3) \sim \exp(-2\pi\Omega^2/\dot{\nu}), \quad (7)$$

where $\dot{\nu}$ is the frequency sweep rate. Therefore, assuming $\Omega^2 > \dot{\nu}$, the spin population initially in branch 1 remains unchanged throughout the LZ crossing. The situation is different, however, for the spin population in branch 2, whose jump probability to branch 3 is approximately given by

$$p(2|3) \sim \exp\left\{-2\pi\left(\frac{\Omega(G+A_{zx})}{4(\omega_{0I}+A_{zz}+F)}\right)^2/\dot{\nu}\right\}, \quad (8)$$

where $F$ and $G$ are functions of the relative orientation of the magnetic field $(\theta,\phi)$, and we are assuming $\Omega^2 > \dot{\nu}$, see SI, Section II. The LZ dynamics in this case is partially non-adiabatic meaning that the spin population initially in branch 2 bifurcates to create a net nuclear spin population difference (Fig. 1C). More generally, the condition for the generation of nuclear spin polarization during a sweep can be formally stated as $p(2|3) > p(1|3)$. We later show the carbon polarization in our simplified NV–$^{13}$C model system (~35% in the calculation of Fig. 1D) is consistent with the observed levels of *bulk* $^{13}$C polarization in our samples (typically in the 0.1-0.3% range). Interestingly, we note that for a frequency sweep starting above, not below, the set of avoided crossings, it is the population in branch 3, not in branch 2, the one that bifurcates. Therefore, an adapted analysis shows that the sign of the end $^{13}$C polarization — calculated as a fractional population difference, see SI, Section II — depends on the direction of the frequency sweep, i.e., a low-to-high sweep yields positive nuclear magnetization whereas the opposite is true for a high-to-low sweep (Fig. 1D).

A comparison with experiment is presented in Fig. 1E, where we probe the bulk $^{13}$C polarization induced at 38 mT under 532 nm illumination; inductive $^{13}$C detection upon MW and optical excitation is carried out at high field with the help of a 7 T NMR system adapted with a sample shuttling device (25). In this particular case, we use a single diamond crystal oriented so that all four NV orientations form the same angle with the applied magnetic field, and limit the MW sweep to a range around the $m_S = 0 \leftrightarrow m_S = -1$ subset of transitions. Consistent with theory (Fig. 1D), we find that reversing the sign of the frequency sweep yields a $^{13}$C NMR signal of opposite phase, indicative of polarization inversion. Note that the phase in the $^{13}$C NMR spectrum attained upon a low-to-high MW sweep coincides with that observed in the thermal signal at 7 T (acquired without optical excitation and/or sample shuttling), hence lifting the ambiguity in the *absolute* sign of the measured nuclear spin polarization.

The structure of the LZ crossings is, in fact, more complex than the one presented in Fig. 1B (corresponding to a comparatively weak hyperfine coupling). Fig. 2 shows the typical energy diagrams for $\|A\|\sim 500$ kHz and 4 MHz (left and right plot sets, respectively), both near the $m_S = 0 \leftrightarrow m_S = -1$ and $m_S = 0 \leftrightarrow m_S = +1$ set of transitions (upper and lower rows, respectively). Comparing Figs. 2A and 2B ($m_S = -1$ manifold), we find that despite the growing frequency separation between the LZ crossings, the asymmetry in the gap size (and thus in the population transfer between branches) is not lifted, i.e., net nuclear spin polarization of the same (sweep-direction-dependent) sign is generated in all cases (even if the efficiency changes with hyperfine coupling strength and relative magnetic field alignment, see below).

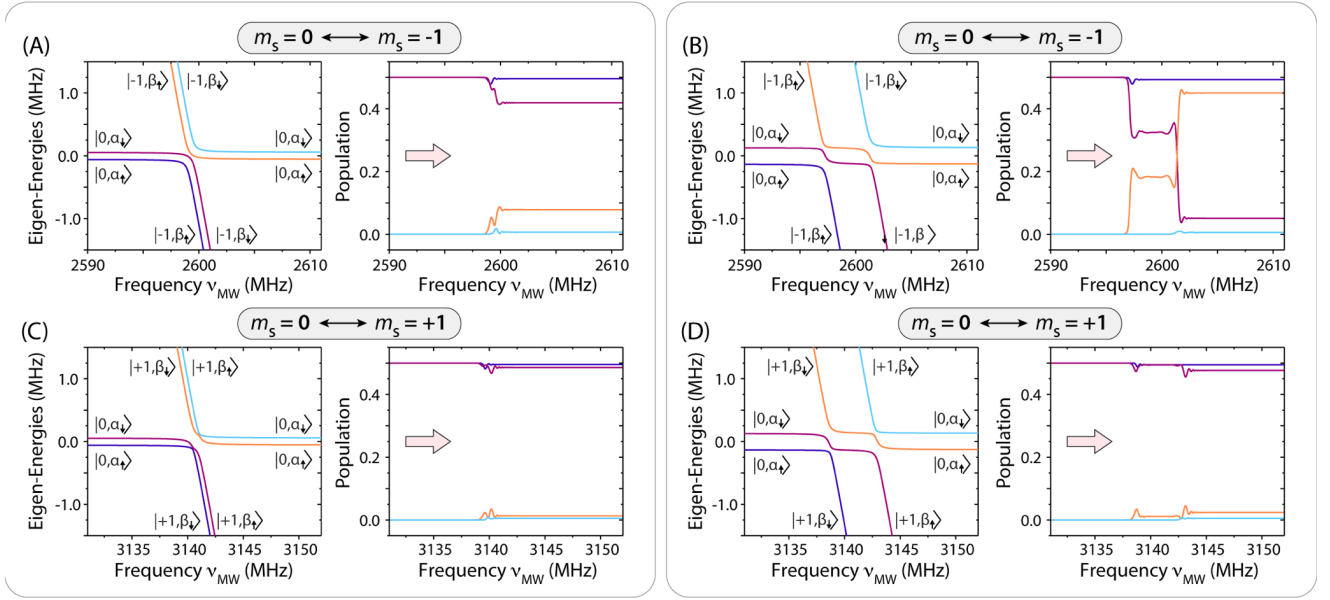

**Fig. 2**. Manifold-dependent nuclear spin polarization dynamics. (A) Eigenenergies of $H_{eff}$ as a function of the MW frequency $\nu_{MW}$ near the $m_S = 0 \leftrightarrow m_S = -1$ subset of transitions and evolution of populations upon a single low-to-high frequency sweep in the presence of a weak hyperfine coupling (right and left, respectively). (B) Same as in (A) but for a stronger hyperfine coupling where all transitions can be individually resolved. (C, D) Same as in (A, B) for the $m_S = 0 \leftrightarrow m_S = +1$ subset of transitions. In (A) through (D), the color code for the right plot follows the notation introduced in the left plot. In (A) and (C), $A_{zz} = A_{xx} = A_{yy} = 500$ kHz, in (B) and (D), $A_{zz} = A_{xx} = A_{yy} = 4$ MHz. In all cases, $A_{zx} = 0.3 A_{zz}$, $\Omega = 250$ kHz, and $\theta = 30°, \phi = 60°$.

By contrast, a quick inspection of Figs. 2C and 2D shows that the energy level structure near the crossings — and corresponding polarization yield — is markedly different in the $m_S = +1$ manifold. Here, reversal in the order of the 'allowed' and 'forbidden' transitions traversed during a sweep (respectively connecting branches with the same 'up' or 'down' nuclear spin character) makes the generation of $^{13}$C polarization inefficient. The impact of the order reversal can be better visualized in Fig. 2D, where the greater hyperfine coupling leads to four resolved LZ crossings. Assuming initialization into $m_S = 0$ and a low-to-high frequency sweep, the large gap in the first avoided crossing — proportional to $\Omega$ — makes this passage predominantly adiabatic. Correspondingly, the subsequent population exchange during the second (narrower) crossing becomes ineffective in creating net nuclear polarization, as the probability of finding the NV–$^{13}$C system in either branch still amounts to approximately 50% (see right plot in Fig. 2D); a similar reasoning applied to the ensuing pair of crossings in the present example confirms that no net polarization can emerge from a sweep of the $m_S = 0 \leftrightarrow m_S = +1$ subset. Note, however, that because in the present low-field regime $\|\mathbf{A}\| > \gamma_C B$, the order in the transitions during a sweep depends on the sign of $A_{zz}$, which can be positive or negative with nearly equal probability. Therefore, there is no intrinsic difference in the polarization efficiency associated to the $m_S = -1$ or $+1$ manifolds, as the dynamics reverses upon an overall sign change of the hyperfine coupling constants. In other words, low-to-high (high-to-low) MW sweeps across either subset in a bulk crystal should yield net positive (negative) $^{13}$C spin polarization. We return to this point later.

To gain a fuller understanding of the dynamics underlying the generation of *bulk* $^{13}$C magnetization we investigate the polarization efficiency as a function of the hyperfine coupling. In Fig. 3A we spin initialize the NV electronic spin to about 5% and determine the steady-state polarization $P_C$ of the coupled $^{13}$C spin as we repeat the MW frequency sweep multiple times; this strategy more closely reproduces our experimental conditions (see Methods and SI). We attain comparable nuclear spin polarization for hyperfine couplings $\|\mathbf{A}\| \gtrsim 500$ kHz and up to 10 MHz. As expected, the efficiency of the polarization transfer process decays for weaker couplings although care must be exercised when correlating the end polarization of a particular $^{13}$C nucleus and its impact on the observed bulk NMR signal. Specifically, as the hyperfine interaction weakens, the number $N$ of carbon spins featuring a lower level of coupling increases (nearly quadratically). Further, weaker couplings considerably facilitate nuclear spin flip-flops between neighbors and hence are instrumental in enabling the generation of bulk nuclear polarization. This is shown in Fig. 3B where we plot the $^{13}$C spin energy splitting $\delta$ within the $m_S = 0$ manifold assuming a 10 mT field: Except for NVs perfectly aligned with $B$, hyperfine contributions (stemming from the last term in Eq. (3)) can quickly dominate over the Zeeman term, thus leading to a hyperfine-dependent frequency mismatch between carbons. This effect is only moderate when $\|A\| \lesssim 1$ MHz, suggesting that carbons featuring weak to moderate couplings have a comparatively larger influence on the observed $^{13}$C NMR signal.

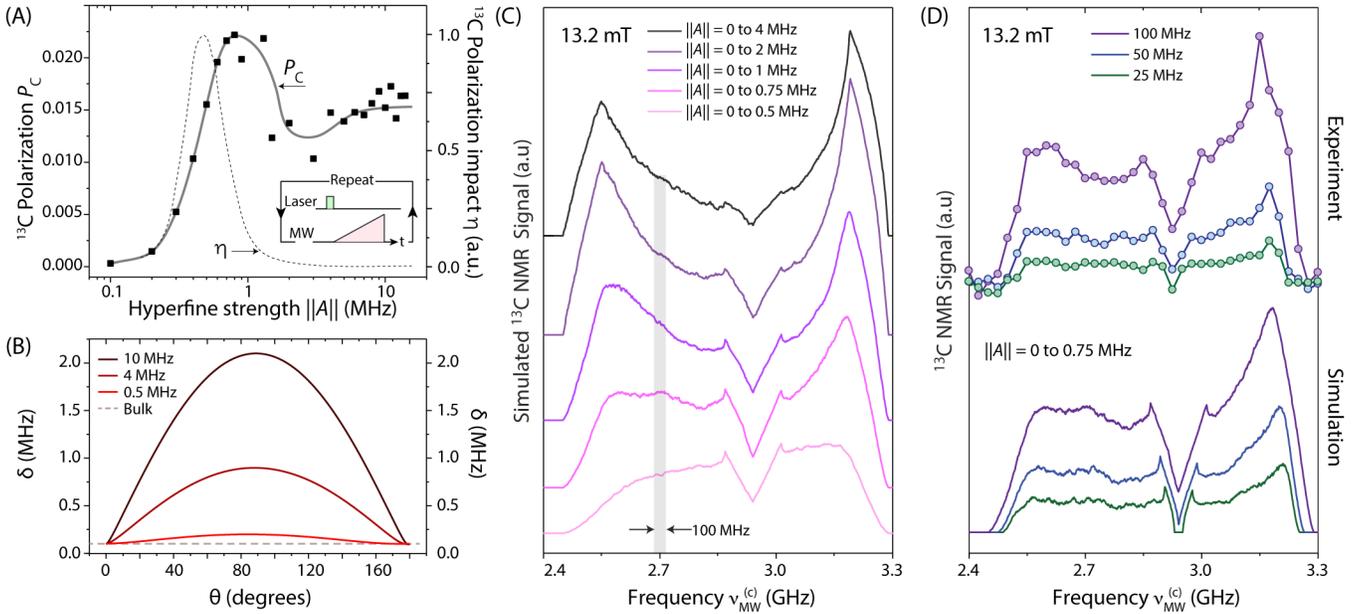

**Fig. 3.** 'Strong' vs. 'moderate' hyperfine couplings. (A) Calculated steady-state $^{13}$C spin polarization after repeated low-to-high frequency sweeps across the $m_S = 0 \leftrightarrow m_S = +1$ subset for a variable hyperfine coupling $A_{zz}$; the initial NV spin polarization is 5%, the sweep rate is 40 MHz/ms, the Rabi field is 250 kHz, and we assume $A_{zz} = A_{xx} = A_{yy}$ and $A_{zx} = 0.3 A_{zz}$; the dots indicate a representative polarization for a given relative orientation of the magnetic field $B$ and the solid line is a guide to the eye. The dashed line indicates the estimated relative contribution $\eta$ to the observed $^{13}$C NMR signal amplitude (right vertical axis). (B) Calculated energy splitting $\delta$ between the NV–$^{13}$C eigenenergies within the subspace $m_S = 0$ as a function of $\theta$ for $\phi = 0$. (C) Modeled $^{13}$C spin NMR signal for a powdered diamond sample and a 100-MHz-wide MW frequency sweep centered at a variable frequency $\nu_{MW}^{(c)}$. In all calculations, we assume the external magnetic field is $B = 13.2$ mT and consider hyperfine interactions over the range $0 - \|A_{max}\|$. (D) Same as in (C) but for a variable sweep range, as calculated numerically or observed experimentally (lower and upper trace sets, respectively). For the calculated plots, we use $\|A_{max}\| = 750$ kHz and all averages emerge over $1.5 \times 10^4$ configurations; throughout the experiments, the number of MW sweeps per data point is $10^3$, and the number of repeats is 30, with all other conditions remaining as in Fig. 1. In (C-D) we displace all traces horizontally for clarity.

We can qualitatively gauge the influence of a given NV–$^{13}$C pair with hyperfine coupling strength $\|A\|$ in generating the observed NMR signal through the 'impact factor' $\eta(\|A\|) \propto \int_0^\pi d\theta \sin^2\theta \int_0^{2\pi} d\phi \, N P_C g(\delta)$ where $N \propto 1/\|A\|^2$, and $g(\delta(\theta,\phi,\|A\|))$ is a function that quantifies the ability of a $^{13}$C spin to spin diffuse with its neighbors. Though the latter is presently unknown, we get a crude estimate by assuming $g(\delta) \propto \exp(-(\delta - \gamma_C B/2\pi)^2/2\|A\|_{max}^2)$ where $\|A\|_{max}$ represents a critical hyperfine coupling, of order ~1 MHz (see below). The result — shown as a dashed line in Fig. 3A — suggests that a very selective set of moderately coupled carbons — featuring $\|A\|$ between 0.5 and 0.8 MHz — is effective in producing the observed bulk NMR signal, very much in contrast with prior studies (6, 11) where strongly coupled carbons (~10 MHz and up) were seen to be dominant.

We confirm the ideas above through the observations in Figs. 3C and 3D: Here, we consider NV–$^{13}$C pairs extracted from a statistical distribution of hyperfine couplings, and calculate for each pair the $^{13}$C spin polarization after a 100 MHz local sweep of variable central frequency $\nu_{MW}^{(c)}$; throughout these calculations we assume the relative orientations of the magnetic field and NV axis are random (see SI, Section III). Fig. 3C shows the results for distributions of hyperfine couplings with a variable upper threshold $\|A\|_{max}$, ranging from 4 MHz down to 0.5 MHz (respectively ordered from top to bottom in the figure). Notably, the calculated responses display a marked asymmetry between the sides of the spectrum associated to transitions involving the $m_S = -1$ or $m_S = +1$ NV spin states (lower and higher frequency regions, respectively). Further, the overall spectral shape is sensitive to the range of hyperfine couplings taken into account, progressively evolving from a structured shape peaked at the edges of the spectrum, towards a flatter, more uniform distribution for weaker $\|A\|_{max}$.

Fig. 3D displays two sets of spectra, both experimental and calculated (top and bottom sets, respectively), which we now can use for direct comparison: In the upper set, each circle represents the amplitude of the measured 7 T $^{13}$C NMR signal upon multiple low-to-high-frequency sweeps over a MW band of pre-defined, variable width centered at a variable frequency $\nu_{MW}^{(c)}$. From an inspection of Fig. 3C, we find best agreement with experiment for carbon distributions where $\|A\|_{max} \approx 750$ kHz, hence indicating that strongly-coupled carbons do not significantly contribute to the observed bulk NMR signal. Note that, despite the lineshape changes, the calculated spectra remain consistent with our observations even as we reduce the sweep bandwidth from 100 MHz, to 50 MHz, to 25 MHz (purple, light blue, and green traces, respectively). Further, because contributions to the calculated spectra stemming from carbons with hyperfine coupling lower than ~200 KHz are negligible (see SI, Section III), the observed bulk nuclear spin polarization must be interpreted as mediated by a select shell of moderately coupled nuclei around the NV (i.e., 200 kHz < $\|A\|$ < 750 kHz), consistent with the calculated 'impact factor' $\eta$ in Fig. 3A.

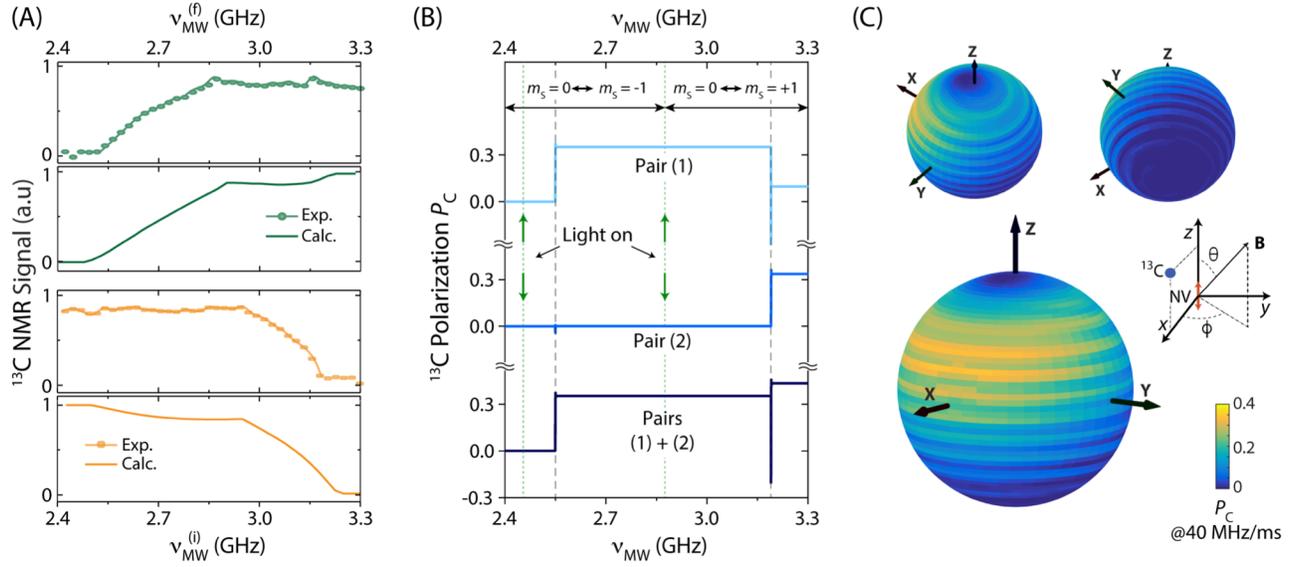

**Fig. 4.** Optimal spin polarization transfer: Frequency and angular dependence. (A) (Top) Measured (full circles) and calculated (solid line) $^{13}$C NMR signal upon multiple frequency sweeps; each sweep goes from a fixed MW start frequency $\nu_{MW}^{(i)} = 2.4$ GHz, to a variable end frequency $\nu_{MW}^{(f)}$. (Bottom) Same as above but for a variable start frequency $\nu_{MW}^{(i)}$ to a fixed final frequency $\nu_{MW}^{(f)} = 3.3$ GHz. In both cases the sweep is from low to high frequencies and thus yields a positive $^{13}$C polarization. All other experimental and modeling conditions as in Fig. 3D. (B) Calculated $^{13}$C spin polarization for two example NV–$^{13}$C pairs featuring similar hyperfine couplings ($\|A\| = 500$ kHz) upon a low-to-high-frequency sweep; the NV–$^{13}$C coupling is assumed positive (negative) for Pair 1 (Pair (2)). Vertical dashed lines indicate the location of the crossings within the $m_S = \pm 1$ manifolds (we are assuming $\theta = 30$ deg). For simplicity, we assume the NV is optically pumped into $m_S = 0$ both at the beginning and midpoint of the sweep (dotted vertical line). (C) Calculated $^{13}$C spin polarization as a function of the orientation of the magnetic field relative to the NV axis, aligned parallel to the z-axis. For these calculations, we use a hyperfine coupling tensor of the form $A_{zz} = A_{xx} = A_{yy} = 0.5$ MHz and $A_{zx} = 0.3 A_{zz}$.

Although the broad spectra in Fig. 3 suggest nearly uniform contributions from NVs in all orientations, the process leading to bulk nuclear spin polarization is considerably more complex. The green circles in Fig. 4A show the result from an experiment where each data point reflects the $^{13}$C NMR signal amplitude upon multiple low-to-high MW frequency sweeps of increasing bandwidth, i.e., the start frequency $\nu_{MW}^{(i)} = 2.4$ GHz remains unchanged while the end frequency $\nu_{MW}^{(f)}$ gradually increases. We find that the NMR signal first grows almost linearly, to subsequently plateau at a maximum once $\nu_{MW}^{(f)} \gtrsim 2.95$ GHz, i.e., once $\nu_{MW}^{(f)}$ reaches the set of transitions involving the $m_S = +1$ NV state; when compared to the thermal signal amplitude, this maximum signal corresponds to a $^{13}$C spin polarization of order 0.2%. We find a similar (though complementary) result if we set the final MW frequency to $\nu_{MW}^{(f)} = 3.3$ GHz and gradually change the start point in the sweep $\nu_{MW}^{(i)}$ towards lower frequencies (orange circles). These observations — qualitatively reproduced by our model, see solid traces in Fig. 4A — can be interpreted in terms of a partial polarization cancellation during the sweep. Fig. 4B illustrates this process through a representative example in which we study the effect of a low-to-high frequency sweep on two individual NV–$^{13}$C pairs featuring hyperfine interactions of the same strength ($\|A\| = 500$ kHz) but of opposite signs. In agreement with the results in Fig. 2, the carbon with a positive (negative) coupling polarizes positively upon crossing the set of transitions involving the $m_S = -1$ ($m_S = +1$) NV spin state. Interestingly, however, the positively coupled carbon loses its polarization when the MW reaches the subset of crossings involving $m_S = +1$, with the result that the net nuclear spin magnetization remains roughly unchanged during the second half of the sweep. In other words, no net increase in the $^{13}$C NMR signal is to be expected when extending the sweep range to include the full set of transitions, as observed experimentally.

We gain additional insight on the NMR signal formation by calculating the steady-state nuclear spin polarization in an individual NV–$^{13}$C pair ($\|A\| = 500$ kHz) for different relative orientations of the magnetic field and NV axes. Despite the broad spectral response observed in Figs. 3C and 3D — naively indicative of angle-insensitive nuclear spin polarization — we find a complex dependence, both in terms of the polar and azimuthal angles (Fig. 4C). Remarkably, our calculations indicate nuclear spin polarization is more efficiently produced in the case of misaligned NVs (i.e., $\theta \neq 0$), which sheds light on why this approach works so effectively in a powdered sample. On the other hand, the polarization transfer process does not show cylindrical symmetry, a reflection of the azimuthal angle dependence in the last term of Eq. (3). Note that although the calculated nuclear polarization is sizable only in the upper hemisphere of the plot, an equivalent response — this time with optimum in the lower hemisphere — can be attained by changing the sign of the hyperfine coupling, from positive (used in the present example) to negative. A crude estimate that takes into account the concentration of NVs and the $^{13}$C spin-lattice relaxation time, shows that the calculated average carbon polarization per sweep in our two-spin model (of order 5%, see Fig. 4C) is consistent with the observed *bulk* $^{13}$C polarization (of order 0.1%, see SI, Section IV).

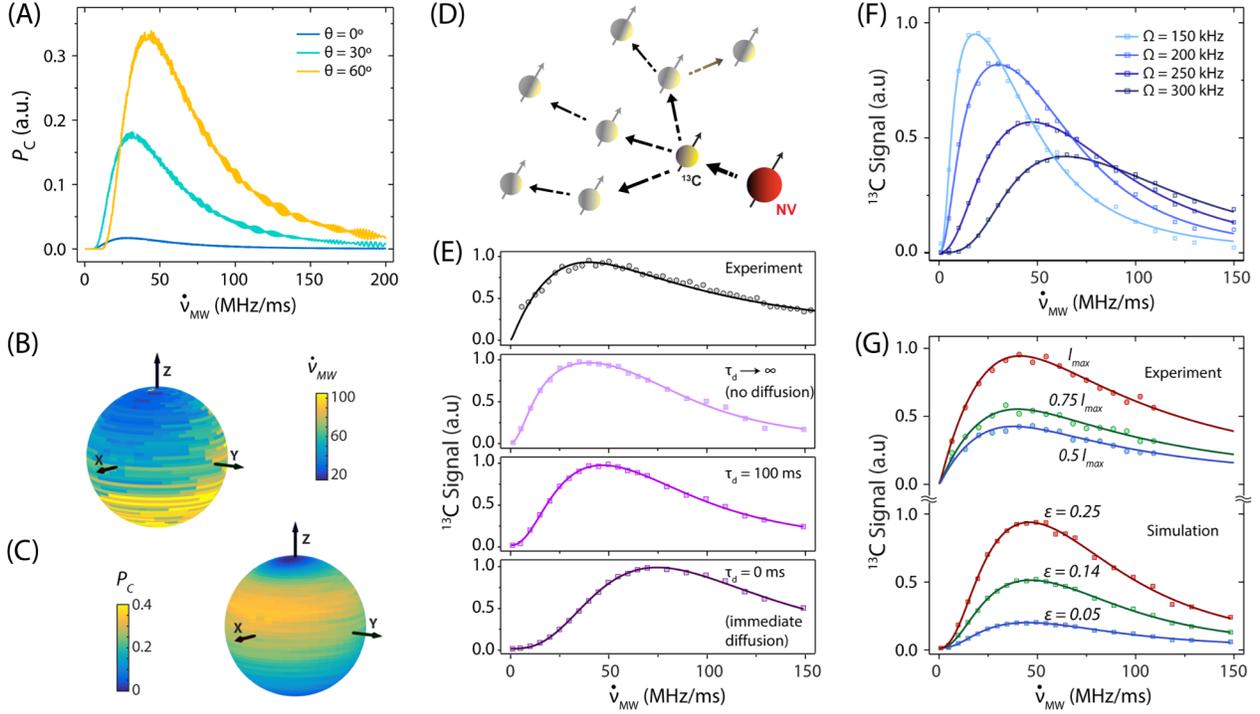

**Fig. 5.** Impact of frequency sweep rate and illumination intensity on the $^{13}$C spin polarization. (A) Calculated $^{13}$C spin polarization upon a single sweep as a function of the MW frequency sweep rate for the case of a moderately-coupled NV–$^{13}$C pair ($\|A\| = 500$ kHz) and different relative orientations of the external magnetic field. For simplicity, we assume the NV has been fully polarized into $m_S = 0$ prior to the sweep and ignore the effect of illumination thereafter; all other conditions as in Fig. 1. (B) Optimal frequency sweep velocity as a function of the magnetic field direction for the $^{13}$C spin considered in (A). (C) Spin polarization for the same $^{13}$C spin as a function of the magnetic field direction assuming in each case the sweep rate is the optimum possible. (D) Nanoscale spin geometry. The polarization from a carbon coupled to an NV spin (yellow and red solid circles, respectively) diffuses via homo-spin couplings with its neighbors (semi-transparent solid circles). (E) The upper plot is the measured $^{13}$C NMR signal amplitude (dark circles) as a function of the sweep rate; the total spin pumping time for each measurement is set to 10 s (see text). The lower traces are the calculated $^{13}$C signals assuming variable spin diffusion time $\tau_d$. (F) Calculated $^{13}$C magnetization for different Rabi field amplitudes as a function of the MW frequency sweep rate. (G) Measured and calculated $^{13}$C NMR signals (upper and lower sets, respectively) at different illumination intensities as a function of the MW sweep rate. In the experimental traces, the maximum laser intensity is $I_{max} = 1$ W. For the calculated traces, we use a fixed Rabi field amplitude $\Omega = 250$ kHz and variable NV spin initialization into $m_S = 0$ quantified via the parameter $\varepsilon \in [0,2]$, see SI, Section IV. All calculations in (E) through (G) assume continuous laser excitation, a magnetic field $B = 13.2$ mT, and a 100 ms spin diffusion time (unless explicitly noted otherwise). Each point emerges from average over all field orientations and hyperfine couplings assuming $\|A\| < 1$ MHz; solid traces are guides to the eye.

Finally, we investigate the impact of the MW frequency sweep rate, set to $\dot{\nu}_{MW} = 40$ MHz/ms in all the results presented thus far. We start by calculating the $^{13}$C spin response in a moderately coupled NV–$^{13}$C pair ($\|A\| = 500$ kHz) for different orientations of the magnetic field. From a quick inspection of Fig. 5A, we conclude that the polarization transfer is most efficient at intermediate sweep rates — of order 40-50 MHz/ms — where non-adiabaticity at the LZ crossings is optimal. Consistent with the observations in Fig. 4C, the absolute $^{13}$C polarization attained at $\dot{\nu}_{MW} = 40$ MHz/ms diminishes for NVs aligned with **B**. On the other hand, the optimal sweep velocity can vary substantially depending on the exact field direction (Fig. 5B), although the absolute change is rather moderate within the set of angles $(\theta, \phi)$ where the polarization transfer is most efficient. In particular, we find that the angular region where the $^{13}$C polarization is maximum remains similar to that in Fig. 4C, even if we choose for each $B$ field direction the optimal sweep rate (Fig. 5C).

Since the time required to complete multiple sweeps inherently depends on the chosen sweep rate, a comparison with experimental observations as a function of $\dot{\nu}_{MW}$ must necessarily take into account nuclear spin diffusion from the target carbon into the bath (Fig. 5D). While a full quantum mechanical description is difficult, we resort to a phenomenological approach where spin diffusion takes the form of a pseudo 'spin-lattice relaxation' process affecting the polarization of the carbon directly coupled to the NV (26); our code, however, keeps track of the total magnetization injected into the nuclear spin system so as to make the end result proportional to the observed $^{13}$C NMR signal (see SI, Section IV).

Experimental plots of the dynamically pumped $^{13}$C NMR signal amplitude as a function of the MW sweep rate are presented in the upper trace of Fig. 5E; to maximize the polarization buildup, the total pumping time at each point is kept constant at $T_{1C} = 10$ s (coincident with the $^{13}$C spin-lattice relaxation time at $B \sim 13.2$ mT), meaning that $n$, the number of sweeps, gradually changes depending on the specific value of $\dot{v}_{MW}$ according to the relation $n = T_{1C}\dot{v}_{MW}/\Delta v$, where $\Delta v = 370$ MHz is the frequency bandwidth. The lower traces in Fig. 5E show the results of numerical simulations where we incorporate the above conditions; throughout these calculations, each data point emerges from a complex average, namely, we sample the hyperfine coupling from a statistical distribution (similar to that used in Fig. 3D, i.e., $0 \leq \|A\| \leq 750$ kHz), and vary the magnetic field direction over all possible orientations relative to the NV axis. To save time (these calculations are demanding and thus require considerable computing power, see Section IV of the SI) we limit the total illumination time to 1s.

In qualitative agreement with our experiments, we find a sharp initial growth followed by a slower decay at faster sweep rates, with an optimum around 40-50 MHz/ms. Taking spin diffusion into account (lower traces in Fig. 5E) shifts the optimum sweep rate to greater values because the larger number of sweeps per unit time effectively enhances the total nuclear magnetization produced. It would be premature, however, to elaborate on the spin diffusion dynamics at play, as the overall shape of the response is sensitive to the Rabi field amplitude (Fig. 5F). Indeed, this latter dependence may be responsible for the main differences between theory and experiment, as the fixed direction of the MW field in the lab frame amounts to a random orientation relative to the NV axis, hence leading to a variable effective MW amplitude.

Interestingly, greater laser powers lead to greater $^{13}$C NMR signals without distorting the overall shape of the response (upper data set in Fig. 5G). This observation — correctly reproduced by our model, see lower data set in Fig. 5G — is a consequence of the relatively mild illumination we employ herein (1 W laser focused over an 4-mm-diameter spot). Under these conditions, the probability of optically exciting an NV precisely during a subset of LZ crossings is relatively low — particularly if, as shown above, $\|A\| \lesssim 750$ kHz, see Fig. 2 — meaning that the polarization transfer takes place coherently and NV repolarization preferentially occurs between successive crossings. More intense laser powers, therefore, lead to better NV spin initialization and, consequently, to enhanced $^{13}$C polarization, as observed experimentally. We caution, however, this picture breaks down for strongly-coupled carbons — where LZ crossings split into a resolved series, see Figs. 2B and 2D — because optical re-initialization of the NV between crossings within the same $m_S = -1$ or $m_S = +1$ manifold typically causes depolarization (see SI, Section V).

**Conclusions**

Continuous optical illumination accompanied by repeated MW frequency sweeps leads to efficient spin polarization transfer from NVs to neighboring carbons in powdered diamond at 10-30 mT. Our observations can be reproduced via an average Hamiltonian describing the effective rotating-frame interaction between the NV and a neighboring $^{13}$C spin. In this picture, spin transfer takes place via a Landau-Zener-like process where nuclear polarization emerges as a consequence of the asymmetry in the adiabaticity parameter characterizing avoided crossings between branches with the same or opposite nuclear spin character; the polarization sign depends on the direction of the sweep, whereas its level relates to the hyperfine coupling, the misalignment, and the particular set of crossings involved. From comparison with $^{13}$C-NMR-detected NV spin spectra, we conclude that the polarization transfer to bulk nuclei is mediated by carbons with hyperfine couplings within a narrow range, $200$ kHz $\lesssim \|A\| \lesssim 750$ kHz. Although comparable levels of nuclear polarization can be attained virtually at all NV spin frequencies, the polarization transfer is sensitive to the relative direction of the magnetic field, with the optimum occurring for carbons associated to misaligned NVs. Further, competing nuclear spin polarization and de-polarization processes limit the range of the frequency sweep necessary to attain maximum $^{13}$C NMR signals to approximately half the NV spin resonance spectrum. The NMR signal response as a function of the frequency sweep rate shows an optimum around 40-50 MHz/ms, consistent with the (calculated) values required to optimize the spin transfer during the LZ crossings.

Our findings open interesting opportunities for further optimization as well as for fundamental and applied work. For example, the angular dependence of the transfer on the $B$ field axis — particularly the azimuthal dependence, see Fig. 4C — suggests that additional NMR signal gain could be attained by making the field direction undergo a suitable time evolution. Also to consider is the sweep rate, which, rather than constant, could be gradually incremented with growing frequencies so as to match the optimum observed at different orientations (Fig. 5B). Along the same lines, another possibility is to use several MW sources to generate multiple sweeps running simultaneously but shifted in frequency; NMR signal amplification is expected when the time separation between successive sweeps — all sharing the same frequency sweep rate — are brought to a minimum defined by the NV repolarization time (27).

Additional work will be needed to investigate the impact of other experimental parameters such as the illumination conditions, and the influence of paramagnetic defects other than the NV. For example, at the relatively mild laser intensities used herein — of order 100 mW/mm$^2$ — the NV polarization rate amounts to ~100 Hz implying that the steady-state NV spin polarization, reaching at best 10%, is far below the optimum. On the other hand, light-induced decoherence during the LZ crossings must ultimately hinder the spin transfer process, which suggests there must be an optimum illumination intensity. Whether or not this regime can be reached without complications from NV photo-ionization (or the ionization of other impurities) is a question that can only be addressed through subsequent studies over a larger range of laser powers and using illumination wavelengths other than 532 nm.

Another pending question is the response as a function of the magnetic field amplitude, here constrained to less than 20 mT. Extending the present studies to greater fields — particularly those above 100 mT — is an attractive route to prolong the $^{13}$C spin-lattice relaxation time — here limited by cross-relaxation with paramagnetic nitrogen impurities (25) — and hence potentially augment the end polarization. Initial observations at ~100 mT showed no enhancement, though a more systematic study — ideally encompassing greater fields — is in order. Several complications, both experimental and theoretical, must be overcome to accomplish this task. Among them is the limited frequency bandwidth typical in most MW sources and amplification systems, normally circumvented at high fields (e.g., $B \sim 300$ mT) through the use of tuned, narrow-band MW cavities and variable magnetic fields; it is not clear, however, this strategy shares the same flexibility as the approach pursued herein. On the theoretical side, additional work will be necessary to extend the present formalism — valid in the limit where the impact of the magnetic field can be treated perturbatively — to the regime where the Zeeman interaction becomes dominant over the NV crystal field. Though some key ingredients remain unchanged (12), this regime is expected to differ from the present one in important ways, including the type of carbons mediating the polarization transfer to the bulk and, most notably, the impact of misalignment on the NV spin initialization. These studies must also encompass the case where the NV spin simultaneously interacts with more than one carbon nucleus, inherent to $^{13}$C-enriched samples and thus important in applications where diamond particles serve as an imaging contrast agent or as the source of hyperpolarization for target fluids.

**Materials and Methods**

Most experiments presented in this paper are carried out using E6 diamond particles with average size of ~200 μm and NV concentration of 1 ppm; more recent work, however, has attained virtually identical results with particle sizes down to 1 μm, either in the form of a dry powder or in solution. The mass of the entire sample of particles is 7.50 ± 0.25 mg. Using the known density of diamond (3.51 mg/mm3), the total sample volume is calculated to be 2.14 ± 0.07 mm$^3$. Dividing by the individual particle volume, the number of diamond particles is found to be 287 ± 27 diamonds. For the experiments in Fig. 1E, we employ a 3×3×0.3 mm$^3$ type 1b single crystal diamond oriented so that all four NV directions form a 54.7 deg. angle with the applied magnetic field.

To induce and detect dynamic nuclear polarization, we use a custom-built sample shuttling device. In a typical experiment, simultaneous MW excitation and 1W, 4-mm-diameter optical illumination at 532 nm take place at a variable low magnetic field (5-36 mT), upon which the sample quickly moves to the sweet spot of a 7T NMR magnet for inspection; the NMR signal amplitude is extracted from the height of the $^{13}$C spectrum obtained upon Fourier transform of the free-induction decay (FID) resulting from resonant single-pulse radio-frequency excitation at 7 T. We refer the reader to Ref. (25) for further details.

Our numerical calculations are conducted using a Matlab code developed in house (see SI, Section IV). We consider individual NV–$^{13}$C pairs whose dynamics follows the effective rotating-frame Hamiltonian presented in Eq. (4); we model the impact of light through sudden projections into the $m_S = 0$ subspace. We sample over all relative orientations of the magnetic field, and, when necessary, over all hyperfine couplings using suitable statistical distributions. To reduce the computing time, we resort to the facilities of the CUNY High Performance Computing Center. We refer the reader to Section IV the Supplementary Information for additional details.

ACKNOWLEDGEMENTS. P.R.Z., S.D., and C.A.M. acknowledge support from the National Science Foundation through grant NSF-1401632, and from Research Corporation for Science Advancement through a FRED Award; they also acknowledge access to the facilities and research infrastructure of the NSF CREST IDEALS, grant number NSF-HRD-1547830. All authors acknowledge the CUNY High Performance Computing Center (HPCC). The CUNY HPCC is operated by the College of Staten Island and funded, in part, by grants from the City of New York, State of New York, CUNY Research Foundation, and National Science Foundation Grants CNS-0958379, CNS-0855217 and ACI 1126113.

# Supporting Information

**Dynamics of frequency-swept nuclear spin optical pumping in powdered diamond at low magnetic fields**

Pablo R. Zangara[1], Siddharth Dhomkar[1], Ashok Ajoy[3], Kristina Liu[3], Raffi Nazaryan[3], Daniela Pagliero[1], Dieter Suter[5], Jeffrey A. Reimer[4], Alexander Pines[3], Carlos A. Meriles[1,2]

[1]Dept. of Physics, CUNY-City College of New York, New York, NY 10031, USA.
[2]CUNY-Graduate Center, New York, NY 10016, USA.
[3]Department of Chemistry, University of California Berkeley, and Materials Science Division Lawrence Berkeley National Laboratory, Berkeley, California 94720, USA.
[4]Department of Chemical and Biomolecular Engineering, and Materials Science Division Lawrence Berkeley National Laboratory University of California, Berkeley, California 94720, USA.
[5]Fakultät Physik, Technische Universität Dortmund, D-44221 Dortmund, Germany.

## I. The effective secular approximation

Here we show how to derive Eq. (4) of the main text from the application of average Hamiltonian theory (AHT) to Eq. (1), using the energy constrains in Eq. (2). We start by writing the Hamiltonian in Eq. (1) as

$$H = \Delta((S^z)^2 - S(S+1)/3) - \gamma_e(B_x S^x + B_y S^y + B_z S^z) - \gamma_C \mathbf{B} \cdot \mathbf{I} + A_{xx} S^x I^x$$
$$+ A_{yy} S^y I^y + A_{zz} S^z I^z + A_{zx}(S^x I^z + S^z I^x). \quad (A.1)$$

Following the standard AHT recipe (1,2), we split (A.1) into two parts, $H = H_\Delta + H_1$, where $H_\Delta = \Delta((S^z)^2 - S(S+1)/3)$ encloses the highest energy scale (or, equivalently, provides for the fastest dynamics), and $H_1$ is automatically defined as $H - H_\Delta$. The zeroth order in AHT is given by

$$\bar{H}^{(0)} = \frac{1}{T_\Delta} \int_0^{T_\Delta} U_\Delta^\dagger(t') H_1 U_\Delta(t') dt', \quad (A.2)$$

where $T_\Delta = 2\pi/\Delta$ and $U_\Delta(t) = \exp\left\{-\frac{i}{\hbar} H_\Delta t\right\}$. It is useful to define $\mathcal{H}_1(t) = U_\Delta^\dagger(t) H_1 U_\Delta(t)$, which we rewrite as,

$$\mathcal{H}_1(t) = -\gamma_e(B_x S_t^x + B_y S_t^y(t) + B_z S^z) - \gamma_C \mathbf{B} \cdot \mathbf{I} + A_{xx} S_t^x I^x + A_{yy} S_t^y I^y + A_{zz} S^z I^z +$$
$$+ A_{zx}(S_t^x I^z + S^z I^x), \quad (A.3)$$

where

$$S_t^x = U_\Delta^\dagger(t) S^x U_\Delta(t) = \frac{\sqrt{2}}{2} \begin{bmatrix} 0 & \exp\left\{\frac{i}{\hbar}\Delta t\right\} & 0 \\ \exp\left\{-\frac{i}{\hbar}\Delta t\right\} & 0 & \exp\left\{-\frac{i}{\hbar}\Delta t\right\} \\ 0 & \exp\left\{\frac{i}{\hbar}\Delta t\right\} & 0 \end{bmatrix}, \quad (A.4)$$

$$S_t^y = U_\Delta^\dagger(t) S^y U_\Delta(t) = \frac{\sqrt{2}i}{2} \begin{bmatrix} 0 & -\exp\left\{\frac{i}{\hbar}\Delta t\right\} & 0 \\ \exp\left\{-\frac{i}{\hbar}\Delta t\right\} & 0 & -\exp\left\{-\frac{i}{\hbar}\Delta t\right\} \\ 0 & \exp\left\{\frac{i}{\hbar}\Delta t\right\} & 0 \end{bmatrix}. \quad (A.5)$$

Upon integrating, it is straightforward to show that

$$\bar{H}^{(0)} = -\gamma_e B_z S^z - \gamma_C \mathbf{B} \cdot \mathbf{I} + A_{zz} S^z I^z + A_{zx} S^z I^x, \tag{A.6}$$

which is, in fact, a valid secular approximation in the aligned case ($\theta = 0$).

The following order in AHT is given by

$$\bar{H}^{(1)} = \frac{-i}{2T_\Delta} \int_0^{T_\Delta} dt_1 \int_0^{t_1} dt_2 [\mathcal{H}_1(t_1), \mathcal{H}_1(t_2)]. \tag{A.7}$$

In what follows, we disregard any term in the previous commutator involving $-\gamma_C \mathbf{B} \cdot \mathbf{I}$ since it produces negligible contributions. Any term in the commutator $[\mathcal{H}_1(t_1), \mathcal{H}_1(t_2)]$ would then be of the form $[S^z, S_t^\alpha]$ or $[S_t^\alpha, S_{t'}^{\alpha'}]$, where $\alpha, \alpha' \in \{x, y\}$ and $t, t' \in \{t_1, t_2\}$. Let us start by considering the terms of the form $[S^z, S_t^\alpha]$:

$$[S^z, S_t^x] = i S_t^y, \tag{A.8}$$

$$[S^z, S_t^y] = -i S_t^x. \tag{A.9}$$

Then,

$$\frac{-i}{2T_\Delta} \int_0^{T_\Delta} dt_1 \int_0^{t_1} dt_2 S_{t_2}^x = \frac{\sqrt{2}}{4\Delta} \begin{bmatrix} 0 & 1 & 0 \\ -1 & 0 & -1 \\ 0 & 1 & 0 \end{bmatrix}, \tag{A.10}$$

$$\frac{-i}{2T_\Delta} \int_0^{T_\Delta} dt_1 \int_0^{t_1} dt_2 S_{t_2}^y = \frac{\sqrt{2}i}{4\Delta} \begin{bmatrix} 0 & -1 & 0 \\ -1 & 0 & 1 \\ 0 & 1 & 0 \end{bmatrix}, \tag{A.11}$$

$$\frac{-i}{2T_\Delta} \int_0^{T_\Delta} dt_1 \int_0^{t_1} dt_2 S_{t_1}^x = \frac{\sqrt{2}}{4\Delta} \begin{bmatrix} 0 & -1 & 0 \\ 1 & 0 & 1 \\ 0 & -1 & 0 \end{bmatrix}, \tag{A.12}$$

$$\frac{-i}{2T_\Delta} \int_0^{T_\Delta} dt_1 \int_0^{t_1} dt_2 S_{t_1}^y = \frac{\sqrt{2}i}{4\Delta} \begin{bmatrix} 0 & 1 & 0 \\ 1 & 0 & -1 \\ 0 & -1 & 0 \end{bmatrix}. \tag{A.13}$$

These terms induce transitions between the subspace $m_S = 0$ and the subspaces $m_S = \pm 1$. The matrix element for these transitions scale as $(\gamma_e B)^2/\Delta \sim 30\text{MHz} \ll \Delta$ so, they are suppressed by the zero-field splitting induced by the crystalline field and we can safely neglect them.

Let us consider the terms $[S_t^\alpha, S_{t'}^{\alpha'}]$ with $\alpha \neq \alpha'$. After some algebra,

$$\frac{-i}{2T_\Delta} \int_0^{T_\Delta} dt_1 \int_0^{t_1} dt_2 [S_{t_1}^x, S_{t_2}^y] = \frac{i}{2\Delta} \begin{bmatrix} 0 & 0 & -1 \\ 0 & 0 & 0 \\ 1 & 0 & 0 \end{bmatrix}, \tag{A.14}$$

$$\frac{-i}{2T_\Delta} \int_0^{T_\Delta} dt_1 \int_0^{t_1} dt_2 [S_{t_1}^y, S_{t_2}^x] = \frac{i}{2\Delta} \begin{bmatrix} 0 & 0 & -1 \\ 0 & 0 & 0 \\ 1 & 0 & 0 \end{bmatrix}. \tag{A.15}$$

These contributions induce transitions between the subspaces $m_S = +1$ and $m_S = -1$, with a matrix element scaling, at best, as $(\gamma_e B)^2/\Delta \sim 30\text{MHz} < \gamma_e B$. Thus, even if not as strong as before, there is a truncation of at least one order of magnitude due to the electron Zeeman splitting. We therefore disregard these terms.

Finally, we consider the terms $[S_t^\alpha, S_{t'}^\alpha]$:

$$\frac{-i}{2T_\Delta}\int_0^{T_\Delta} dt_1 \int_0^{t_1} dt_2 [S_{t_1}^x, S_{t_2}^x] = \frac{1}{2\Delta}\begin{bmatrix} 1 & 0 & 1 \\ 0 & -2 & 0 \\ 1 & 0 & 1 \end{bmatrix}, \tag{A.16}$$

$$\frac{-i}{2T_\Delta}\int_0^{T_\Delta} dt_1 \int_0^{t_1} dt_2 [S_{t_1}^y, S_{t_2}^y] = \frac{1}{2\Delta}\begin{bmatrix} 1 & 0 & -1 \\ 0 & -2 & 0 \\ -1 & 0 & 1 \end{bmatrix}. \tag{A.17}$$

By the same criteria, the off-diagonal elements have to be disregarded as they represent transitions between the subspaces $m_S = +1$ and $m_S = -1$, which are truncated by the Zeeman splitting. However, the diagonal matrix elements are in fact relevant, since they can induce transitions between different $^{13}$C spin states.

Now we insert Eq. (A.3) into Eq. (A.7) and trace the origin of the commutators $[S_t^\alpha, S_{t'}^\alpha]$. These can arise either from terms like $-(\gamma_e B_\alpha)^2 [S_{t_1}^\alpha, S_{t_2}^\alpha]$ or terms like $-\gamma_e B_\alpha A_{\alpha\alpha} I^\alpha [S_{t_1}^\alpha, S_{t_2}^\alpha]$. In the first case, the diagonal matrix elements in Eqs. (A.16) and (A.17) produce energy shifts two orders of magnitude smaller than the actual gap $\sim\Delta$ and can be ignored. The second case stands for actual coupling terms between the $^{13}$C spin states. These terms have matrix elements as large as $\delta \sim \gamma_e B_\alpha A_{\alpha\alpha}/\Delta \sim 1\times 10^{-1}$MHz for $A_{\alpha\alpha} \sim 1$MHz, and up to $\delta \sim 1$MHz for $A_{\alpha\alpha} \sim 10$MHz. Thus, these transitions are critical for the $m_S = 0$ subspace, where the Zeeman splitting $-\gamma_C B$ is no longer the dominant energy scale.

We are left with the following second-order AHT correction according:

$$\bar{H}^{(1)} = \frac{-\gamma_e}{\Delta}[B_x(A_{xx}I^x + A_{zx}I^z) + B_y A_{yy}I^y]\begin{bmatrix} 1 & 0 & 0 \\ 0 & -2 & 0 \\ 0 & 0 & 1 \end{bmatrix}. \tag{A.18}$$

This Hamiltonian represents a secularization of terms not included in $\bar{H}^{(0)}$, and ultimately leads to the desired secular Hamiltonian, i.e. Eq. (3) in main text, $H_{\text{sec}} = H_\Delta + \bar{H}^{(0)} + \bar{H}^{(1)}$. The effective Hamiltonian introduced in Eq. (4) of the main text follows from a rotating frame transformation,

$$H_{\text{eff}} = H_{\text{sec}} - \omega(S^z)^2 + \Omega S^x, \tag{A.19}$$

where $\Omega$ is the Rabi frequency. Note that the term proportional to $(S^z)^2$ — valid only in the limit $\gamma_e B < \Delta$ considered herein — simultaneously takes into account contributions from the rotating and counter-rotating terms stemming from the linearly polarized MW field; depending on the value of $\omega$, one or the other becomes resonant across the set of transitions $(m_S = 0 \leftrightarrow m_S = -1)$ and $(m_S = 0 \leftrightarrow m_S = +1)$. To simplify the notation, we set $\hbar = 1$ throughout our calculations.

## II. Eigenstates and observables

In Fig. S1 we compare the first four eigenstates $|0, \alpha_\uparrow\rangle, |0, \alpha_\downarrow\rangle, |-1, \beta_\uparrow\rangle$, and $|-1, \beta_\downarrow\rangle$ obtained from the exact Hamiltonian (A.1) and the secular approximation $H_{\text{sec}}$ (we omit an equivalent analysis for the subspace $m_s = +1$). The comparison is based on the decomposition of each eigenstate in terms of the computational basis $\{|m_S, m_I\rangle\}$ as a function of the angle $\theta$ (fixed $\phi = 0$). It is worth noting that the states $|0, \alpha_\uparrow\rangle$ and $|0, \alpha_\downarrow\rangle$ remain predominantly given by $|0, \uparrow\rangle$ and $|0, \downarrow\rangle$ respectively in the range $0° \leq \theta \leq 90°$. This is also true (and even more accurately so) for $|-1, \beta_\uparrow\rangle$, and $|-1, \beta_\downarrow\rangle$,

which are essentially given by $|-1,\uparrow\rangle$ and $|-1,\downarrow\rangle$ respectively. The inversion observed at 90° simply corresponds to the change in the preferred direction of quantization $z \to -z$.

As discussed in the main text, the LZ crossings create population imbalances between these states, leading to the observed $^{13}$C polarization. However, even though the whole process takes place at low magnetic field and in presence of light, the actual signal acquisition throughout our experiments is performed by shuttling the sample into a high-field NMR system. Since such a transfer is intrinsically adiabatic, the population imbalance created at low-field remains unchanged in the high-field condition. This suggests a way to define the $^{13}$C polarization in our simulations by algebraically adding the populations in the instantaneous eigen-basis with the sign given by the character of the state (up or down). Alternatively, one can compute the expectation value of the $I_z$ operator rotated in the direction of the magnetic field. In this case, however, the time-dependence of the $^{13}$C polarization exhibits fast coherent oscillations which unnecessarily complicate the numerics. Nevertheless, it is important to stress that both ways of computing the $^{13}$C polarization give essentially the same results, and that both are in agreement with our experimental observations.

We now turn our attention to estimating the energy gaps $\delta E_{13}$ and $\delta E_{23}$ presented in Fig. 1B and the LZ probabilities $p(1|3)$ and $p(2|3)$. We start by assuming the effective Hamiltonian in the aligned case $\theta = 0, \phi = 0$,

$$H_{eff}^{[\theta=0,\phi=0]} = \Delta(S^z)^2 - \gamma_e B_z S^z - \gamma_C B_z I^z + A_{zz}S^z I^z + A_{zx}S^z I^x - \omega(S^z)^2 + \Omega S^x. \tag{A.20}$$

In what follows, we consider the Hilbert subspace spanned by the basis states $\{|0,\uparrow\rangle, |0,\downarrow\rangle, |-1,\uparrow\rangle, |-1,\downarrow\rangle\}$. By introducing the notation $\omega_{0S} = \Delta - |\gamma_e|B_z$ and $\omega_{0I} = \gamma_C B_z$, the matrix representation of $H_{eff}^{[\theta=0,\phi=0]}$ in this subspace is:

$$H_{eff}^{[\theta=0,\phi=0]} = \begin{array}{c} \\ \langle 0,\uparrow| \\ \langle 0,\downarrow| \\ \langle -1,\uparrow| \\ \langle -1,\downarrow| \end{array} \begin{pmatrix} |0,\uparrow\rangle & |0,\downarrow\rangle & |-1,\uparrow\rangle & |-1,\downarrow\rangle \\ -\frac{\omega_{0I}}{2} & 0 & \frac{\Omega}{2} & 0 \\ 0 & \frac{\omega_{0I}}{2} & 0 & \frac{\Omega}{2} \\ \frac{\Omega}{2} & 0 & \omega_{0S} - \frac{\omega_{0I}}{2} - \omega - \frac{A_{zz}}{2} & -\frac{A_{zx}}{2} \\ 0 & \frac{\Omega}{2} & -\frac{A_{zx}}{2} & \omega_{0S} + \frac{\omega_{0I}}{2} - \omega + \frac{A_{zz}}{2} \end{pmatrix}. \tag{A.21}$$

If the MW irradiation is close to the $|0,\downarrow\rangle \leftrightarrow |-1,\uparrow\rangle$ resonance then the two states are degenerate, which means that

$$\omega_{0S} - \frac{\omega_{0I}}{2} - \omega - \frac{A_{zz}}{2} \approx \frac{\omega_{0I}}{2}, \tag{A.22}$$

or, equivalently,

$$\omega_{0S} - \omega_{0I} - \frac{A_{zz}}{2} \approx \omega. \tag{A.23}$$

We therefore rewrite the Hamiltonian as

$$H_{eff}^{[\theta=0,\phi=0]} = \begin{array}{c} \\ \langle 0,\uparrow| \\ \langle 0,\downarrow| \\ \langle -1,\uparrow| \\ \langle -1,\downarrow| \end{array} \begin{pmatrix} |0,\uparrow\rangle & |0,\downarrow\rangle & |-1,\uparrow\rangle & |-1,\downarrow\rangle \\ -\frac{\omega_{0I}}{2} & 0 & \frac{\Omega}{2} & 0 \\ 0 & \frac{\omega_{0I}}{2} & 0 & \frac{\Omega}{2} \\ \frac{\Omega}{2} & 0 & \frac{\omega_{0I}}{2} & -\frac{A_{zx}}{2} \\ 0 & \frac{\Omega}{2} & -\frac{A_{zx}}{2} & \frac{3}{2}\omega_{0I}+A_{zz} \end{pmatrix}. \quad (A.24)$$

The interaction matrix element $\langle 0,\uparrow|H_{eff}^{[\theta=0,\phi=0]}|-1,\uparrow\rangle = \Omega/2$ corresponds to the NV spin flip produced by the Rabi oscillation. As expected, this corresponds to having $\delta E_{13} \sim \Omega$, and accordingly $p(1|3) \sim exp\{-2\pi\Omega^2/\dot{v}_{MW}\}$. The origin of the gap $\delta E_{23}$ is subtler, and in order to provide for an estimate, we first assume $\omega_{0I} \ll A_{zz}$ and focus on the subspace spanned by the states $\{|0,\uparrow\rangle,|0,\downarrow\rangle,|-1,\uparrow\rangle\}$. We incorporate then energy shifts based on second order perturbation theory,

$$H_{eff,reduced}^{[\theta=0,\phi=0]} = \begin{array}{c} \\ \langle 0,\uparrow| \\ \langle 0,\downarrow| \\ \langle -1,\uparrow| \end{array} \begin{pmatrix} |0,\uparrow\rangle & |0,\downarrow\rangle & |-1,\uparrow\rangle \\ -\frac{\omega_{0I}}{2} & 0 & \frac{\Omega}{2} \\ 0 & \frac{\omega_{0I}}{2}-\frac{\Omega^2}{4(\omega_{0I}+A_{zz})} & 0 \\ \frac{\Omega}{2} & 0 & \frac{\omega_{0I}}{2}-\frac{A_{zx}^2}{4(\omega_{0I}+A_{zz})} \end{pmatrix}. \quad (A.25)$$

A fairly good approximation for the gap $\delta E_{23}$ (see below for a comparison with the fully numerical solution) can be obtained by diagonalization,

$$\delta E_{23} \approx \left| \frac{\omega_{0I}}{2} + \frac{(A_{zx}^2-2\Omega^2)}{8(\omega_{0I}+A_{zz})} - \frac{1}{2}\sqrt{\left(\omega_{0I}-\frac{A_{zx}^2}{4(\omega_{0I}+A_{zz})}\right)^2 + \Omega^2} \right| \quad (A.26)$$

Notice, however, that this estimate is nonzero even if $A_{zx} = 0$. This happens because the degeneracy of the states $|0,\downarrow\rangle$ and $|-1,\uparrow\rangle$ (the condition stated in Eq. (A.22)) is broken by the presence of interaction terms with the states $|0,\uparrow\rangle$ and $|-1,\downarrow\rangle$. These interaction terms do contribute to $\delta E_{23}$, but since they are not genuine interaction matrix elements between the states $|0,\downarrow\rangle$ and $|-1,\uparrow\rangle$, they cannot be used to compute the LZ transition probabilities. In other words, the gap $\delta E_{23}$ is not the actual magnitude ruling the LZ process.

The only way to produce a transition between the states $|0,\downarrow\rangle$ and $|-1,\uparrow\rangle$ would be a second order interaction term between them mediated by the intermediate state $|-1,\downarrow\rangle$. The corresponding matrix element for such a virtual interaction (not present in Eq. (A.25)) is given by

$$J_{virtual} = \frac{\langle 0,\downarrow|H_{eff}^{[\theta=0,\phi=0]}|-1,\downarrow\rangle\langle -1,\downarrow|H_{eff}^{[\theta=0,\phi=0]}|-1,\uparrow\rangle}{\frac{\omega_{0I}}{2}-\langle -1,\downarrow|H_{eff}^{[\theta=0,\phi=0]}|-1,\downarrow\rangle}$$

$$J_{virtual} = \frac{-\Omega A_{zx}}{4\left(\frac{1}{2}\omega_{0I}-\frac{3}{2}\omega_{0I}-A_{zz}\right)} = \frac{\Omega A_{zx}}{4(\omega_{0I}+A_{zz})}. \quad (A.27)$$

This provides for a fair estimate of the LZ transition probability at the energy crossing between the branches of states $|0,\downarrow\rangle$ and $|-1,\uparrow\rangle$,

$$p(2|3) \sim exp\left\{-2\pi\left(\frac{\Omega A_{zx}}{4(\omega_{0I}+A_{zz})}\right)^2/\dot{v}_{MW}\right\}. \quad (A.28)$$

Similar arguments can be used for the resonance $|0,\uparrow\rangle \leftrightarrow |-1,\downarrow\rangle$, obtaining the same estimates.

The general case, i.e. arbitrary $(\theta,\phi)$, is more involved. Considering the same 4-state subspace as above, we have

$$H_{eff} = \begin{array}{c c} & \begin{array}{cccc} |0,\uparrow\rangle & |0,\downarrow\rangle & |-1,\uparrow\rangle & |-1,\downarrow\rangle \end{array} \\ \begin{array}{c} \langle 0,\uparrow| \\ \langle 0,\downarrow| \\ \langle -1,\uparrow| \\ \langle -1,\downarrow| \end{array} & \left( \begin{array}{cccc} \frac{-\omega_{0I}}{2}+F & G & \frac{\Omega}{2} & 0 \\ G^\dagger & \frac{\omega_{0I}}{2}-F & 0 & \frac{\Omega}{2} \\ \frac{\Omega}{2} & 0 & \omega_{0S}-\frac{\omega_{0I}}{2}-\omega-\frac{A_{zz}}{2}-\frac{F}{2} & -\left(\frac{G+A_{zx}}{2}\right) \\ 0 & \frac{\Omega}{2} & \left(\frac{G^\dagger+A_{zx}}{2}\right) & \omega_{0S}+\frac{\omega_{0I}}{2}-\omega+\frac{A_{zz}}{2}+\frac{F}{2} \end{array} \right) \end{array}, \quad (A.29)$$

where $F = (\gamma_e/\Delta)B_x A_{zx}$, $G = (\gamma_e/\Delta) \times (B_x A_{xx} - iB_y A_{yy})$, and we are neglecting the mixing created by the terms $-\gamma_C B_x I^x - \gamma_C B_y I^y$. As before, we first assume the MW frequency is in near resonance with the $|0,\downarrow\rangle \leftrightarrow |-1,\uparrow\rangle$ transition, meaning that

$$\omega_{0S} - \frac{\omega_{0I}}{2} - \omega - \frac{A_{zz}}{2} - \frac{F}{2} \approx \frac{\omega_{0I}}{2} - F, \quad (A.30)$$

or, equivalently,

$$\omega_{0S} - \omega_{0I} - \frac{A_{zz}}{2} + \frac{F}{2} \approx \omega. \quad (A.31)$$

Then,

$$H_{eff} = \begin{array}{c c} & \begin{array}{cccc} |0,\uparrow\rangle & |0,\downarrow\rangle & |-1,\uparrow\rangle & |-1,\downarrow\rangle \end{array} \\ \begin{array}{c} \langle 0,\uparrow| \\ \langle 0,\downarrow| \\ \langle -1,\uparrow| \\ \langle -1,\downarrow| \end{array} & \left( \begin{array}{cccc} \frac{-\omega_{0I}}{2}+F & G & \frac{\Omega}{2} & 0 \\ G^\dagger & \frac{\omega_{0I}}{2}-F & 0 & \frac{\Omega}{2} \\ \frac{\Omega}{2} & 0 & \frac{\omega_{0I}}{2}-F & -\left(\frac{G+A_{zx}}{2}\right) \\ 0 & \frac{\Omega}{2} & -\left(\frac{G^\dagger+A_{zx}}{2}\right) & \frac{3\omega_{0I}}{2}+A_{zz} \end{array} \right) \end{array}. \quad (A.32)$$

Note that while the direct matrix element $\langle 0,\uparrow|H_{eff}|-1,\uparrow\rangle = \Omega/2$ still provides for the estimate $\delta E_{13} \sim \Omega$ and $p(1|3) \sim \exp\{-2\pi\Omega^2/\dot{v}_{MW}\}$ remains valid, it is not straightforward to calculate or estimate $\delta E_{23}$. Regardless, we show in Fig. S2 that the estimate for the aligned case (Eq. (A.26)) is still a fair estimate of $\delta E_{23}$ for a large range of values of $\theta$.

As before, we are interested in an interaction matrix element between the states $|0,\downarrow\rangle$ and $|-1,\uparrow\rangle$. Again, this is given by a second order interaction term mediated by the intermediate state $|-1,\downarrow\rangle$. Thus, the estimate for the LZ transition probability is in this case

$$p(2|3) \sim \exp\left\{-2\pi\left(\frac{\Omega(G+A_{zx})}{4(\omega_{0I}+A_{zz}+F)}\right)^2 / \dot{v}_{MW}\right\}. \quad (A.33)$$

Notice that this estimate reduces to Eq. (A.28) in the limit $(\gamma_e B/\Delta) \to 0$.

The results above provide a simple framework to describe the generation of $^{13}$C polarization as a function of the sweep velocity. Indeed, a crude approximation for the nuclear spin polarization can be written as the product $g(\dot{v}_{MW}) \times q(\dot{v}_{MW}) \times (1 - Q(\dot{v}_{MW}))$, where we introduced the notation $Q(\dot{v}_{MW}) = p(1|3)$. The last factor $(1 - Q(\dot{v}_{MW}))$ measures the adiabaticity during the sweep for branch 1. At low-intermediate velocities (where $Q(\dot{v}_{MW}) \sim 0$), the factor $q(\dot{v}_{MW})$ equals the bifurcation probability $p(2|3)$. In the limit of fast sweeps, however, one has to correct $p(2|3)$ with an extra factor $(1 - Q(\dot{v}_{MW}))$ that accounts for the transition from branch 2 to branch 4, so, in general, $q(\dot{v}_{MW}) \sim p(2|3)(1 - Q(\dot{v}_{MW}))$. Note that a transition between branches 2 and 4 does not generate net polarization and must satisfy the condition $p(2|4) \equiv p(1|3)$. In addition, the correction ensures that the sum of all populations before and after the crossing remains unchanged. Finally, the factor $g(\dot{v}_{MW}) \sim 1 - \exp(\dot{v}_{MW}/k)$ accounts for the cumulative effect of a varying number of sweeps within a fixed measurement time at a given sweep rate, and $k$ is a parameter gauging the impact of spin diffusion. A more detailed discussion on this last point is addressed in Section IV.

## III. Statistical sampling

In the cases where averaging over configurations is required (e.g., Figs. 3, 4A, 5), we perform a simultaneous sampling over angular coordinates $(\theta, \phi)$ and the hyperfine interaction. In particular, we recall that the variables $(\theta, \phi)$ correspond to the direction of the external magnetic field in the crystal-frame where the $z$-direction is given by the NV crystalline field. Thus, we use the standard homogeneous spherical distribution,

$$\theta = \cos^{-1}(2r_1 - 1)$$

$$\phi = 2\pi r_2$$

where $r_1, r_2$ are uniform pseudorandom numbers in the interval (0,1).

The hyperfine tensor is assumed to have the following structure:

$$\mathbf{A} = \begin{bmatrix} A_{xx} & 0 & A_{zx} \\ 0 & A_{yy} & 0 \\ A_{xz} & 0 & A_{zz} \end{bmatrix}, \tag{A.34}$$

with $A_{xx} = A_{yy} = s_1 a$, $A_{zz} = s_2 a$, and $A_{xz} = A_{zx} = 0.3a$. Here $s_1, s_2$ are pseudorandom binary variables that account for sign randomization (they can be either +1 or -1). For each realization, the value of $a$ is taken from a uniform distribution in the interval $(0, \|A\|_{max})$. As an example, we show in Fig. S3 the simulated $^{13}$C NMR signal obtained for different values of $\|A\|_{max}$.

Since the best agreement with the experimental results is achieved when $\|A\|_{max} < 1$ MHz, it is natural to ask if the interaction can be strictly dipolar. In fact, we have verified that nearly identical results can be obtained by using in our simulations the standard dipole-dipole interaction instead of the generic tensor in Eq. (A.34).

## IV. Numerical simulation

In order to reconstruct the $^{13}$C NMR signal as in Figs. 3, 4A, and 5, we compute the explicit time dependence of each MW sweep. More specifically, the time of each sweep is $\tau_s = \Delta v/\dot{v}_{MW}$, where $\Delta v$ is the frequency window of the sweep and $\dot{v}_{MW}$ the sweep velocity. We divide the frequency window in steps of $\delta v = 100$ Hz and evaluate stepwise the time evolution at each of these frequencies by exact diagonalization of the effective Hamiltonian $H_{\text{eff}}$ (Eq. (4) in the main text). The time $t_{\text{bin}}$ spent at each frequency bin is given by the ratio between the time of the sweep and the number of bins, i.e. $t_{\text{bin}} = \tau_s/(\Delta v/\delta v)$. The final state of each bin is used as the initial state for the following bin.

At the beginning of any sweep, we assume that the initial state of the NV is given by

$$\rho_i^{NV} = \frac{(1+\varepsilon)}{3}|0\rangle\langle 0| + \frac{(1-\varepsilon/2)}{3}(|-1\rangle\langle -1| + |+1\rangle\langle +1|), \tag{A.35}$$

where the parameter $\varepsilon$ controls the NV polarization generated by light-induced spin initialization. In practice, this means that we deterministically project the NV state into $\rho_i^{\text{NV}}$. This projection accounts for the repolarization of the NV while keeping the $^{13}$C spin state unchanged. We emphasize that when the sweep is broad enough to encompass both the $m_S = 0 \leftrightarrow m_S = -1$ and $m_S = 0 \leftrightarrow m_S = +1$ sets of transitions, we assume the light intensity is sufficient to repolarize the NV in the time spent sweeping the MW frequency from one set to the next. The same repolarization is assumed to happen in the case of consecutive sweeps of the same subset. Moreover, in all simulations we assume that the NV repolarizes to the same level (defined by the parameter $\varepsilon$) irrespective of the sweep velocity. This is in fact a crude approximation, since for very high velocities there is only a very short time between successive sweeps (or between the two manifolds) and therefore the NV repolarization is less efficient. As we stated in the main text, the NMR signal enhancement is expected to be optimal when the time separation between successive sweeps is brought to a minimum defined by the NV repolarization time. We discuss the case of an unpolarized NV in Section V.

For given angular coordinates $(\theta, \phi)$, it is crucial to know the exact location of each resonance (i.e. LZ crossings) in the frequency space. This is particularly important in reproducing the actual shape (MW frequency dependence) of the experimental NMR signal. So, even though the dynamics of polarization is evaluated in the rotating frame by means of $H_{\text{eff}}$, the actual location of each signal contribution in the frequency axis is determined by diagonalizing the exact Hamiltonian without MW irradiation (i.e. Eq. A.1). This reshuffling procedure allows for the correct distribution of the LZ processes along the frequency domain.

The appropriate quantification of spin diffusion is relevant when reconstructing the NMR signal after multiple sweeps with a fixed total time $T$. In such case, it is relevant to compare the time $\tau_s$ between two successive sweeps with the time $\tau_d$ at which the polarization diffuses away from the $^{13}$C directly coupled to the NV. On the one hand, in the limit of low velocities $\dot{v}_{\text{MW}} \to 0$, we have $\tau_s \gg \tau_d$, so the generated nuclear spin polarization scales linearly with the number of sweeps $n = T/\tau_s$. In the opposite limit of high velocities $\dot{v}_{\text{MW}} \to \infty$, many sweeps take place until the polarization diffuses away from the directly coupled $^{13}$C. In this latter case, the total polarization is dominated by the efficiency of the transfer between the NV and the $^{13}$C. For any intermediate case, a given number of sweeps is performed until the polarization can actually diffuse away and build up the 'bulk' polarization. We show in Fig. S4 a flow chart that explains the algorithm used in our simulation.

Given the complexity of the many-body problem and the energy mismatch between the $^{13}$C coupled and the rest of the 'bulk' carbons, it is hard to have a fair estimate for $\tau_d$. A lower bound is given by the spin-spin interaction time $T_2$, which for $^{13}$C in naturally enriched samples is ~10 ms. However, the actual diffusion process can be much slower than that, with estimated scale as long as ~50×$T_2$ (3).

Due to the uncertainty in some of the parameters (effective laser power, NV spin-lattice relaxation time and level of spin polarization, effective nuclear spin diffusion time, etc.), a comparison between the calculated $^{13}$C spin polarization in our NV–$^{13}$C model upon a single sweep (~5%, see Figs. 4C, 5A, 5C) and the measured *bulk* carbon polarization (of order 0.1%) is difficult. We can, nonetheless, attain a crude estimate when we note that for a sample with natural $^{13}$C content (~1%) and 1 ppm NV concentration, there are approximately $10^4$ carbons per NV. For the optimal conditions of sweep velocity, it takes ~10 ms to complete one full sweep; therefore, assuming a spin diffusion time of 100 ms, a total of 10 sweeps can bring the polarization of a single carbon on par with that of the NV (~10% for our present experimental conditions). During a 10 s illumination, that corresponds to polarizing 100 carbons to about 10 percent, or 10 fully polarized carbons per NV. Out of the $10^4$ carbons, that corresponds to a bulk $^{13}$C spin polarization of 0.1%, comparable to the measured values.

## V. Light-induced stochastic jumps and the ratchet effect

Other than the deterministic projection into state (A.35), an alternative, more realistic way to introduce light-induced NV repolarization makes use of stochastic quantum jumps (4). Here, one can assume that an initially unpolarized NV undergoes an instantaneous repolarization event with some unit time probability $p_r$, in turn, dependent on the light intensity. In such an event or 'jump', the state of the NV collapses into $|0\rangle\langle 0|$. The actual $^{13}$C spin polarization emerges as the result of an average over a sufficiently large number of stories or trajectories.

In order to illustrate the usefulness of the 'jump' approach, we show in Fig. S5 a sequence of three consecutive sweeps. During sweep one, the NV state is polarized by a first jump at a time when the MW frequency is approximately 2.750 GHz, and nuclear spin polarization emerges upon traversing the LZ crossings (see Figs. S5A through S5C). During sweep two (Figs. S5D through S5F), no early NV spin repolarization occurs and traversing the LZ avoided crossings leads to nuclear *depolarization*. A subsequent jump event repolarizes the NV and the third sweep is able to create net polarization again (Figs. S5G through S5I). This example clearly shows that even when rare events of no-repolarization before sweeping happen (and degrade the signal), the system recovers immediately after the next NV spin repolarization. In this sense, the MW sweep in the presence of light acts as a nuclear spin polarization ratchet.

The jump picture is also useful to show the relative fragility of nuclear spin polarization induced via strong hyperfine interactions. Assuming near optimal sweep rate, no polarization can be created if the jump event occurs in between two consecutive LZ crossings within the same $m_s = +1$ or $m_s = -1$ manifold as shown in Fig. S6 for the LZ subset $m_s = 0 \leftrightarrow m_s = -1$ and $\|A\| = 10$ MHz. Since this 'fragile' region where the mechanism is sensitive to light is as large as $A_{zz}$, strongly coupled carbons are comparatively more sensitive to depolarization than those more moderately coupled (i.e. $\|A\| \lesssim 1$ MHz). Note that this observation adds to the trend already highlighted in Fig. 3A of the main text, already favoring moderately coupled carbons in their ability to transfer polarization to the bulk.

The averaging procedure implicit in the use of the quantum jump picture is independent from (and complementary to) the configurational average described in Section III (where the sampling is carried out over all hyperfine couplings and relative magnetic field orientations). So, even though this approach is physically more accurate, its use is computationally more demanding and thus must be restricted to select cases. Whenever possible, nonetheless, we have verified the equivalence between results obtained using the quantum jump picture and the deterministic initialization of the NV spin.

---

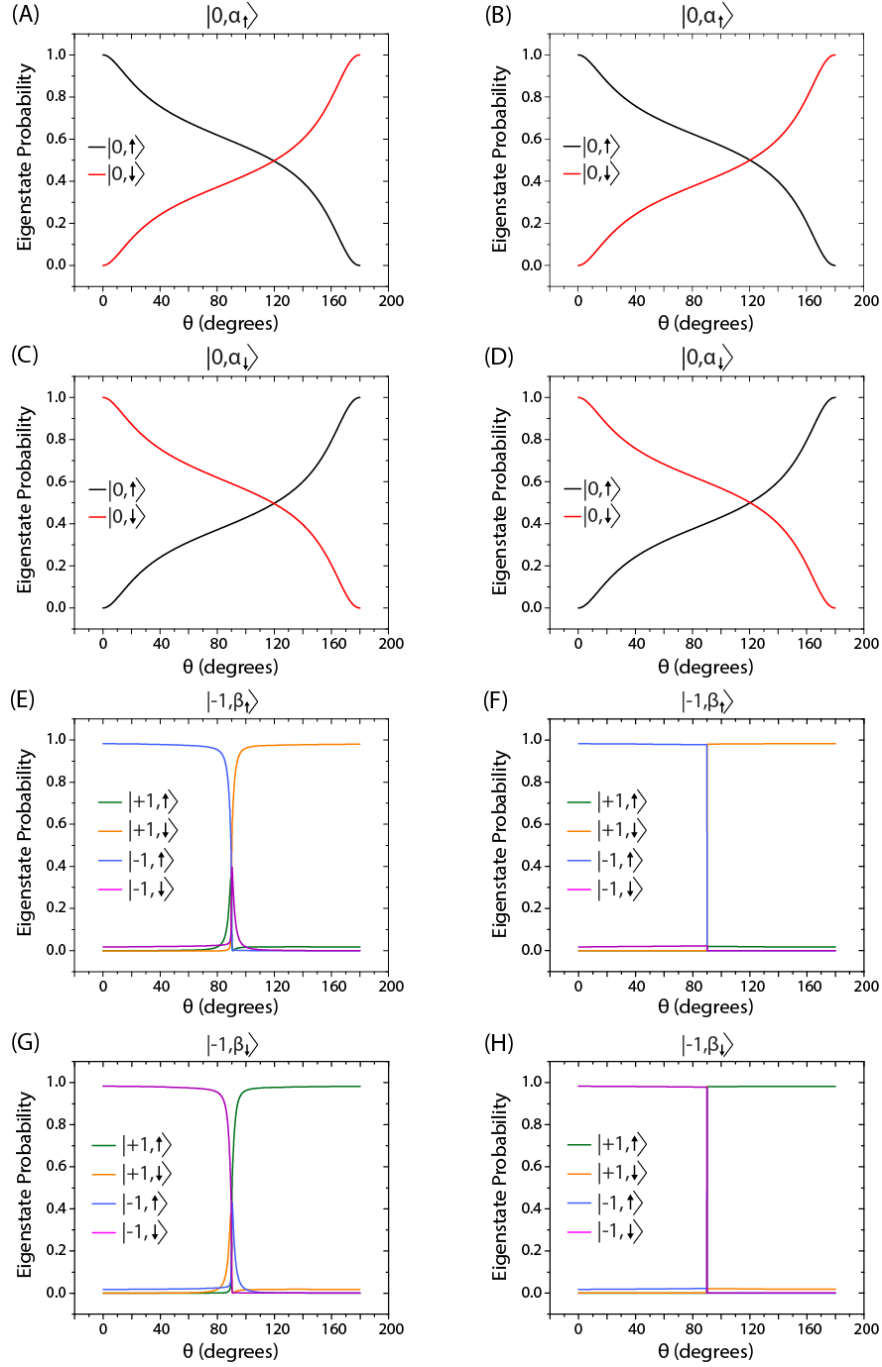

**Figure S1.** Comparison between the exact Hamiltonian $H$ in (A.1) and the secular approximation $H_{\text{sec}} = H_\Delta + \overline{H}^{(0)} + \overline{H}^{(1)}$. In (A), (C), (E) and (G) we plot the decomposition of the exact eigenstates $|0, \alpha_\uparrow\rangle$, $|0, \alpha_\downarrow\rangle$, $|-1, \beta_\uparrow\rangle$ and $|-1, \beta_\downarrow\rangle$ respectively in terms of the computational basis states. We do the same in (B), (D), (F) and (G), but using the eigenstates of $H_{\text{sec}}$. In all cases, $\phi = 0$, $A_{\text{xx}} = A_{\text{yy}} = A_{\text{zz}} = 1$ MHz, $A_{\text{xz}} = 0.3 A_{\text{zz}}$ and $B = 10$ mT.

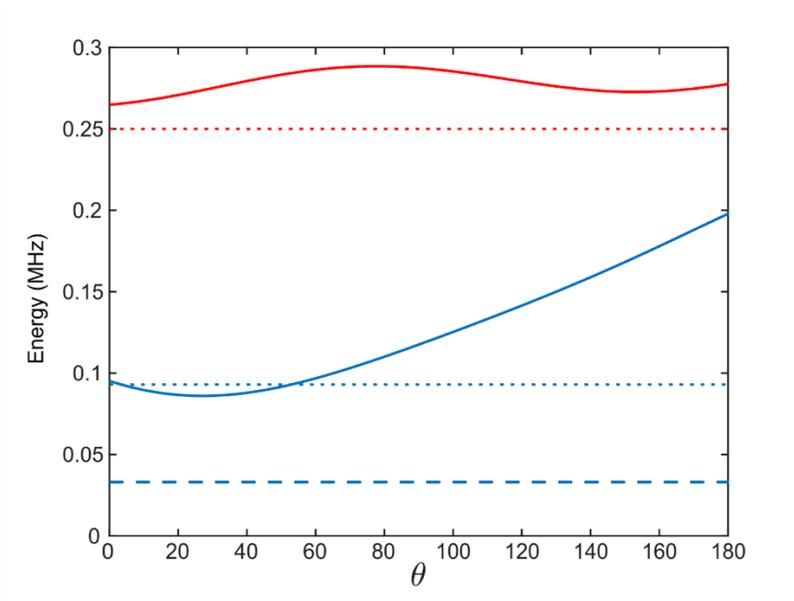

**Figure S2**. Calculated energy gaps $\delta E_{13}$ and $\delta E_{23}$ as a function of $\theta$, for $\phi = 0$ (solid red and blue traces, respectively). The resonance condition corresponds to the transition $|0,\downarrow\rangle \leftrightarrow |-1,\uparrow\rangle$. The choice of parameters is: $A_{zz} = A_{xx} = A_{yy} = 750$ kHz, $A_{zx} = 0.3 A_{zz}$, $|\vec{B}| = 10$ mT, $\Omega = 250$ kHz. The dashed, blue trace corresponds to the "virtual" gap $2J_{virtual} = \Omega A_{zx}/2(\omega_{0I} + A_{zz})$ and the dotted, blue trace corresponds to Eq. (A.26). The dotted, red trace corresponds to the estimate $\delta E_{13} \sim \Omega$.

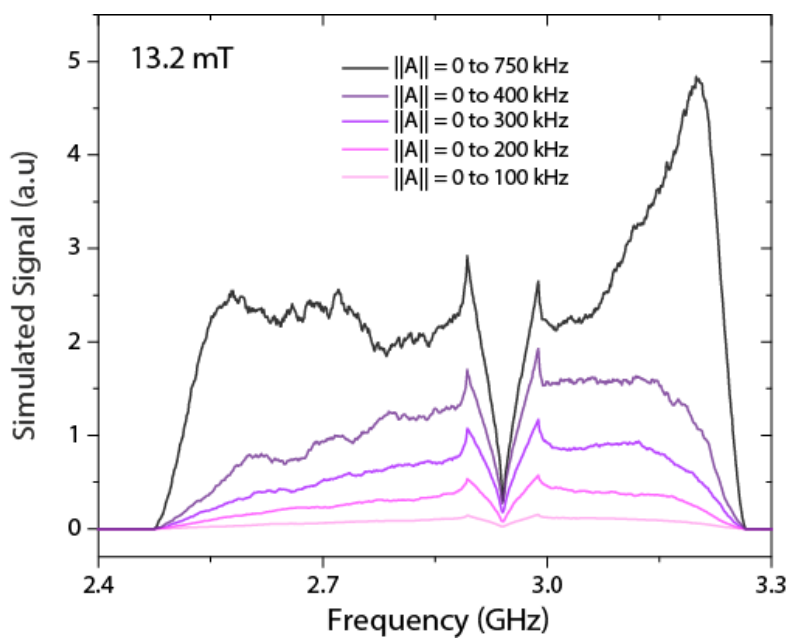

**Figure S3**. Simulated $^{13}$C NMR signal for a 50-MHz-wide MW frequency sweep centered at a variable central frequency. Here, we assume the external magnetic field is $B = 13.2$ mT and consider $1.5 \times 10^4$ configurations for $(\theta, \phi)$ and the hyperfine interaction, whose magnitude is taken from the uniform distribution $[0, \|A\|_{max}]$.

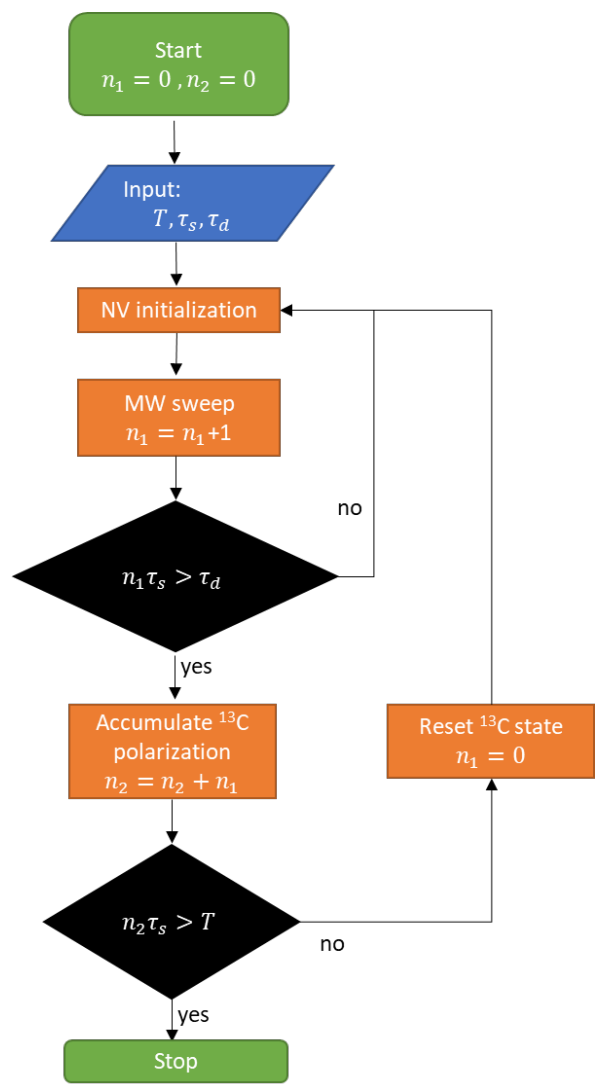

**Figure S4**. Flow chart for simulating the effect of diffusion in a multiple-sweep, fixed-time protocol. The input variables are the total time $T$ (typically 1 sec), the time per sweep $\tau_s = \Delta \nu / \dot{\nu}_{MW}$ and the diffusion time $\tau_d$. The index $n_1$ controls the number of sweeps until a 'diffusion event' takes place. In such a case, the nuclear magnetization is accumulated and the state of the system (NV-$^{13}$C pair) is reset. The algorithm stops after a total time $T$ has elapsed, which means that the index $n_2$ equals $n = T/\tau_s$.

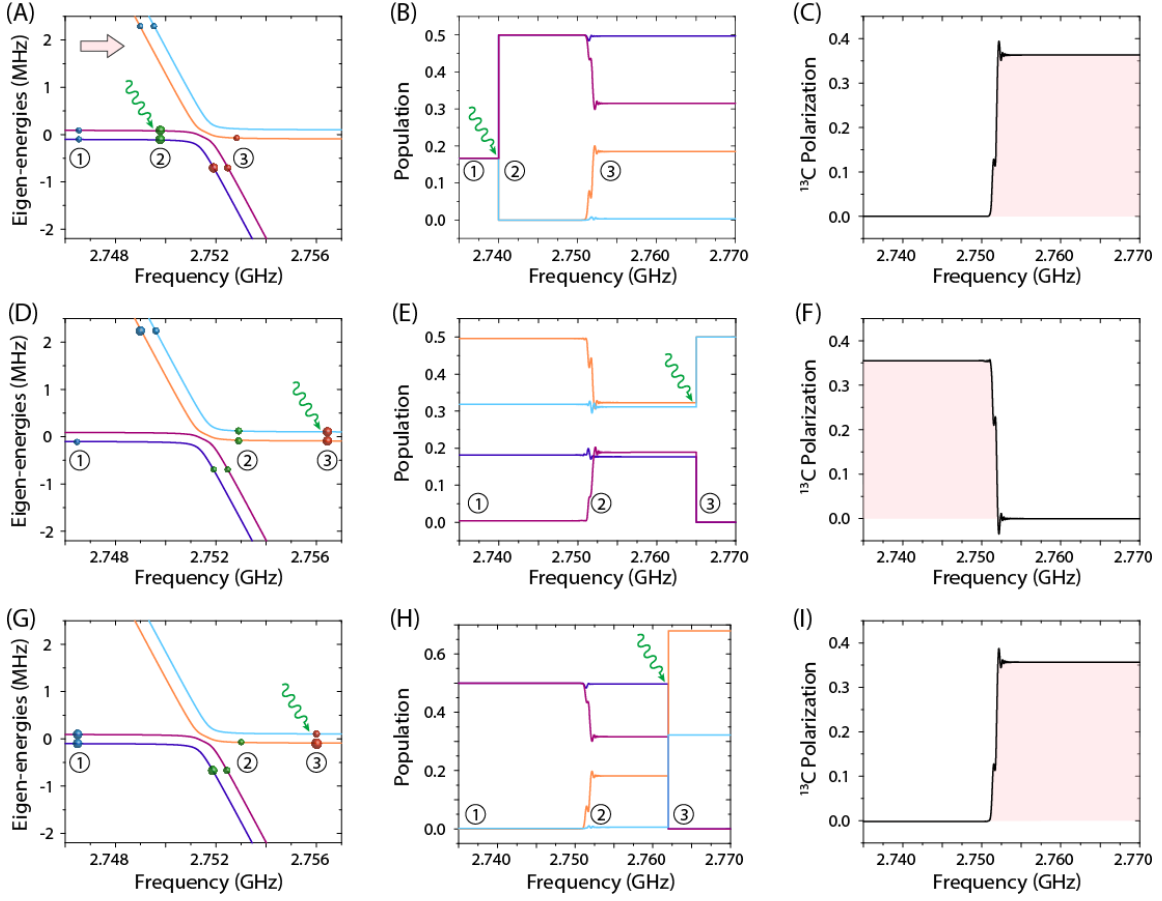

**Figure S5**. Multiple sweeps in the presence of light-induced stochastic jumps. For the present example, $\theta = 65°$, $\phi = 0°$, and $A_{xx} = A_{yy} = A_{zz} = 500$ kHz, and $A_{xz} = 0.3 A_{zz}$. In (A), (D), and (G) the labels for states at each energy curve are the same as in Fig. 2A in the main text. Blue circles indicate initial state ①, green circles indicate the intermediate state ②, and red circles denote the final state ③. A wiggly green arrow indicates a light-induced repolarization event or jump. In (A), we start with a completely unpolarized initial state ①, which subsequently collapses into the subspace $m_s = 0$ upon NV spin optical pumping (state ②); nuclear spin polarization emerges after a MW sweep across the LZ crossing (state ③). In (B) we explicitly show the evolution of these populations and in (C) the corresponding $^{13}$C polarization. In (D-F) we show evolution during the second sweep assuming the initial state ① (blue circles in (D)). After the LZ crossing the polarization is lost (state ②, green circles) since there is no more nuclear spin population imbalance. An NV spin repolarization jump in ③ brings back the NV population to the subspace $m_s = 0$ (with no effect on the $^{13}$C polarization). In (G-I) we show that under these conditions the third sweep creates again $^{13}$C polarization (state ②) and a final NV spin repolarization (state ③) brings the system population to the subspace $m_s = 0$. In this case, a fourth sweep would add more $^{13}$C polarization instead of destroying it.

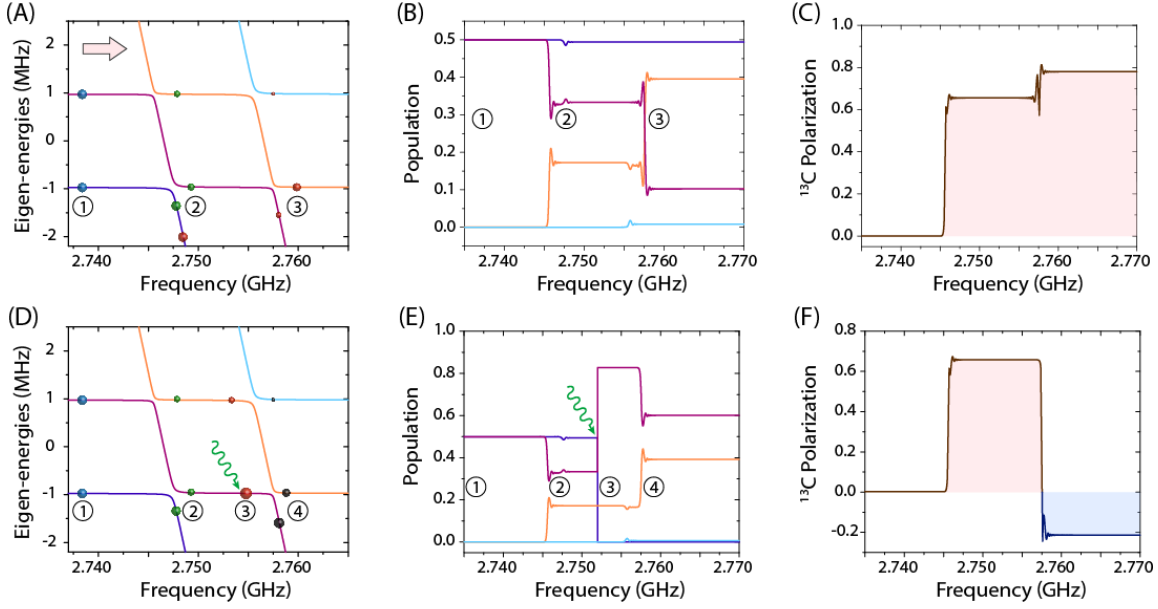

**Figure S6**. Single MW sweeps for large hyperfine in the presence of light-induced stochastic jumps. In this example, $\theta = 65°$, $\phi = 0°$, and $A_{xx} = A_{yy} = A_{zz} = 10$ MHz, and $A_{xz} = 0.3 A_{zz}$. In (A) and (D) the labels for states in each energy curve are the same as in Fig. 2B in the main text. In (A-C) we show a single sweep for an initially polarized NV (blue circles, state ①). The first LZ crossing already generates $^{13}$C polarization (green circles, state ②). The second LZ crossing generates even more nuclear spin imbalance (red circles, state ③). In (D-E) we repeat the same simulation but with a jump at a time between the two LZ crossings. This NV spin repolarization event brings the population in $|-1, \beta_\uparrow\rangle$ back into the state $|0, \alpha_\uparrow\rangle$ (red circles in (D), state ③). The second LZ crossing not only destroys the net $^{13}$C polarization created, but it turns it into negative (black circles, state ④).